\documentclass{aa}  

\usepackage{graphicx}
\usepackage{txfonts}
\usepackage[usenames,dvipsnames]{xcolor}

\makeatletter
\renewcommand*\aa@pageof{, page \thepage{} of \pageref*{LastPage}}
\makeatother

\begin{document}

   \title{The $i$-processes nucleosynthesis during the formation of He-rich hot-subdwarf stars}

   %\subtitle{I. Dependence on how stripped is the star}

   \author{Battich T.
          \inst{1}
          \and
          Miller Bertolami M. M.
          \inst{2,}\inst{3}
          \and
          Weiss A.\inst{1}
          \and
          Dorsch M.\inst{4}
          \and
          Serenelli A. M.
          \inst{5,}\inst{6}
          \and       
          Justham S.\inst{1}
    %      \and
    %      de Mink, S. E.\inst{1} 
          }

   \institute{Max-Planck-Institut f\"ur Astrophysik, Karl-Schwarzschild Strasse 1, D-85748, Garching, Germany\\
              \email{tiara@mpa-garching.mpg.de}
         \and
          Instituto de Astrofísica de La Plata, Consejo Nacional de Investigaciones Científicas y Técnicas Avenida Centenario (Paseo del Bosque) S/N, B1900FWA La Plata.
         \and
          Facultad de Ciencias Astronómicas y Geofísicas, Universidad Nacional de La Plata Avenida Centenario (Paseo del Bosque) S/N, B1900FWA La Plata, Argentina
          \and
          Institut für Physik und Astronomie, Universität Potsdam, Haus 28, Karl-Liebknecht-Str. 24/25, 14476 Potsdam-Golm, Germany
        \and Institute of Space Sciences (ICE, CSIC), Carrer de Can Magrans S/N, E-08193, Cerdanyola del Valles, Spain
          \and Institut d’Estudis Espacials de Catalunya (IEEC), Carrer Esteve Terradas, 1, Edifici RDIT, Campus PMT-UPC, E-08860, Castelldefels,  Spain
             }

   \date{Received too late; accepted even later}

% \abstract{}{}{}{}{}
% 5 {} token are mandatory
 
  \abstract
  % context heading (optional)
  % {} leave it empty if necessary  
   {Intermediate neutron-capture processes are thought to occur in stellar environments where protons are ingested into a hot helium-burning convective region. 
   It has been shown that proton ingestion episodes can happen in the formation of hot-subdwarf stars, and that neutron-capture processes are possible in those cases. Moreover, some helium-rich hot subdwarfs display extraordinarily high abundances of heavy elements such as Zr, Yr and Pb on their surfaces. These elements can be produced by both slow and intermediate neutron-capture processes.}
  % aims heading (mandatory)
   {We explore under which conditions neutron-capture processes can occur in late helium core flashes, i.e. those occurring in the cores of stripped red-giant stars. We explore the dependence of such processes on the metallicity of the star, and on how much the star is stripped before the He-flash takes place.}
  % methods heading (mandatory)
   {We compute evolutionary models through the helium core flash and the subsequent hydrogen ingestion episode in stripped red giant stars. Stellar structure models are then used in post-processing to compute the detailed evolution of neutron-capture elements}
  % results heading (mandatory)
   {We find that for metallicities of $10^{-3}$ and below, neutron densities can be as high as $10^{15}\,$cm$^{-3}$ and intermediate neutron capture processes occur in some of our models. 
   %Moreover, the ashes of hydrogen burning in the He-flash are mixed towards the surface, enriching the surface of the star with heavy elements. 
   The results depend very strongly on the H-envelope mass that survives after the stripping, which alters the nucleosynthesis during the flashes, and therefore the behavior of the convective motions. Interestingly, we find that computed abundances in some of our models closely match the element abundances up to tin observed  for EC 22536-5304, the only well-studied star for which the hot-flasher scenario assumed in our models is the most likely evolutionary path. 
   }
  % conclusions heading (optional), leave it empty if necessary
   {Intermediate neutron capture processes can occur in the He-core flash experienced by the cores of some stripped red giants, and might be connected to the abundances of heavy elements observed in some helium-rich hot-subdwarf stars. The agreement between the observed abundances in EC 22536-5304 and those of our models offers  support to our nucleosynthesis calculations. Moreover, if confirmed, the idea that heavy element abundances retain signatures of the different evolutionary channels opens the possibility that heavy element abundances in iHe-sdOB stars can be used to infer their evolutionary origin.
   }


   \maketitle
%
%________________________________________________________________
\section{Introduction}

Neutron capture processes are responsible for the formation of most of the elements heavier than iron. Traditionally, two main types of neutron-capture processes have been identified: the slow ($s$-) and rapid ($r$-) neutron-capture processes. In the first case, neutron densities are of the order of $N_n \simeq 10^7 - 10^{11}\,$cm$^{-3}$, and the timescale of the neutron captures is longer than that of the $\beta^-$ instability of the neutron-rich nuclei. Therefore, this process proceeds through neutron-rich species close to the valley of stability in the nuclei chart. For much higher neutron densities, $N_n\gtrsim 10^{20}$, the $r$-process takes place, and species further from the valley of stability are formed by successive neutron captures before $\beta^-$ decays lead to the formation of stable nuclei \citep{1957RvMP...29..547B,2020PrPNP.11203766A}. \citet{1977ApJ...212..149C} proposed that neutron-capture nucleosynthesis might happen inside stars at intermediate neutron densities, $N_n \simeq 10^{12}-10^{16}\,$cm$^{-3}$. This process was named the intermediate ($i$-) neutron-capture process. The $i$-process did not receive much attention until some carbon-enhanced metal-poor stars showed abundances of heavy elements that could not be explained by a combination of $r$ and $s$ enrichment (e.g. \citealt{2012AIPC.1484..111L,2015arXiv150505500D}). The $i$-process was also proposed as responsible for the abundances observed in Sakurai's Object \citep{2011ApJ...727...89H}, the central star of a planetary nebula that erupted in 1995 \citep{1996IAUC.6322....1N, 1996ApJ...468L.111D}. The astrophysical sites where the $i$-process can occur as well as its impact on the cosmological chemical evolution are not yet completely understood. One of the main proposed sites is the first thermal pulse of very low metallicity asymptotic giant branch (AGB) stars \citep{2010A&A...522L...6C}. This site has been studied in detail by \cite{2021A&A...648A.119C, 2022A&A...667A.155C, 2021A&A...654A.129G}. Another proposed site being recently discussed are rapidly-accreting white dwarfs, proposed by \cite{2017ApJ...834L..10D} and studied in detail in successive works \citep{2019MNRAS.488.4258D, 2021MNRAS.503.3913D, 2018ApJ...854..105C, 2021MNRAS.504..744S}. There have been other sites studied in the literature, namely, the main helium flash of ultra low metallicity stars in the red giant branch \citep{2010MNRAS.405..177S, 2010A&A...522L...6C, 2013A&A...559A...4C}, low metallicity massive stars \citep{2018ApJ...865..120B,2018MNRAS.474L..37C,2021MNRAS.500.2685C} and very late thermal pulses in post-AGB stars \citep{2011ApJ...727...89H, 2018JPhG...45e5203D}. All of these sites have in common that the $i$-process nucleosynthesis happens when protons are ingested by a convective He-burning zone. In this situation, protons are captured by $^{12}$C nucleus producing $^{13}$N which decays to $^{13}$C in the order of minutes. Because the temperature is high enough for He-burning, $\alpha$ particles can be captured by $^{13}$C producing $^{16}$O and a flux of neutrons. In contrast with the $s$-process happening in thermal pulses of AGB stars, the timescale of burning and mixing in this case is much shorter, and can lead to higher neutron densities that might allow $i$-process nucleosynthesis. 

In addition to the sites for $i$-process nucleosynthesis mentioned above, there is another astrophysical scenario in the evolution of some low-mass stars where a proton ingestion episode (PIE) to a convective He-burning zone can occur. A PIE might happen during the formation of He-rich hot-subdwarf stars (He-sdOBs, \citealt{2001ApJ...562..368B,2003ApJ...582L..43C, 2023A&A...680L..13B}). He-sdOB stars are located at the extreme end of the horizontal branch, with temperatures between $32500$ and $40000\,$K, have masses of about half a solar mass, and the majority of them are believed to be in the core He-burning phase \citep{2016PASP..128h2001H}. Two main scenarios have been proposed for the formation of He-rich hot subdwarfs: the merger of two low-mass white dwarfs (WD, \citealt{Webbink1984WDmergers,2011MNRAS.410..984J,2012MNRAS.419..452Z,2016MNRAS.463.2756H,2018MNRAS.476.5303S,2022MNRAS.511L..60M}), and the occurrence of a core He flash in a stripped red giant when the star has lost almost all of its envelope and had departed from the red-giant branch to higher temperatures, a scenario referred to as `hot He-flash' scenario \citep{1996ApJ...466..359D,2001ApJ...562..368B,2004ApJ...602..342L,2008A&A...491..253M}. This last scenario can happen either through binary interaction, where a red giant loses its envelope after a common envelope or a mass transfer episode \citep{2003MNRAS.341..669H,2018MNRAS.475.4728B,2020A&A...642A..97K}, or in isolated stars if they are born with high He abundances and low metallicity, as can be found in some globular clusters \citep{2017A&A...597A..67A}. Single hot-subdwarf stars could also be formed after a red giant close to the tip of the red giant branch engulfs a substellar companion \citep{2010AIPC.1314...85H,2020MNRAS.496..612H}. 

In the hot He-flash scenario, the core He flash develops once the H-rich envelope has been almost completely removed. If the mass of the H-rich envelope is low enough, the H-burning layer will become inactive. If the He-flash happens under these conditions, the convective region driven by the He-flash can reach the H-rich zone, leading to a PIE and making the star H deficient (e.g. \citealt{2008A&A...491..253M}). Moreover, a group of hot-subdwarf stars with an intermediate enrichment on He (iHe-sdOBs) feature overabundances of heavy elements like strontium, yttrium, zirconium and/or lead of about $10000$ times the solar value \citep{2011MNRAS.412..363N,2013MNRAS.434.1920N,2020MNRAS.491..874N,2017MNRAS.465.3101J,2019MNRAS.489.1481J,2019A&A...629A.148L,2020MNRAS.499.3738O,2019A&A...630A.130D,2020A&A...643A..22D,2021A&A...653A.120D}.  Because hot-subdwarf stars are compact and hot objects with radiative atmospheres, they suffer from diffusion processes without the retarding interference of winds and convection. The high abundances of heavy elements might be due to the action of radiative levitation, pushing heavy elements towards the line-formation regions of the stars \citep{2011MNRAS.412..363N, 2011A&A...529A..60M}. However, in \cite{2023A&A...680L..13B}, we have recently shown that neutron-capture processes can occur in the PIE following a late hot-flash in a stripped red giant. Moreover, the superficial abundance pattern of heavy elements obtained in the model of \cite{2023A&A...680L..13B} agrees qualitatively well with the abundances observed in the iHe-sdOBs. 

It has been shown before that the strength and length of the H-burning phase following a PIE in a late hot-flash model depends upon the intensity of the H-burning shell just before the occurrence of the He-core flash which, in turn, depends on the H-rich envelope mass. Consequently, the development of the PIE depends on how much the red giant has been stripped before undergoing the He flash (see, e.g., \citealt{2008A&A...491..253M, 2018A&A...614A.136B}). The amount of H burned and the time span of the PIE can significantly affect the heavy-metal nucleosynthesis. In \cite{2023A&A...680L..13B} we have explored a model with a mass of $0.4846\,M_{\odot}$ and a H-rich envelope mass of $6.08\times10^{-4}\,M_{\odot}$ at the moment of the He flash. In this work, we expand the study of the evolution of the PIE and the nucleosynthesis of heavy elements depending on the H-rich envelope mass at the moment of the He-flash and for different initial compositions. The paper is organized as follows. In Section \ref{sec:methods} we describe the models, methods and codes used for the calculation of detailed nucleosynthesis. In Section \ref{sec:vlhf} we describe in detail the late core He-flash models focusing on the evolution of the PIEs and the creation of neutrons. In Section \ref{sec:nuc} we describe the results of detailed nucleosynthesis calculations. In Section \ref{sec:discuss} we discuss our results. In Section \ref{sec:conclusions} we summarize our conclusions.

%__________________________________________________________________

\section{Method and codes}\label{sec:methods}

In order to study the development of $i$-process in late He-flash models, we first calculate the evolution of stellar models with the stellar evolution code \texttt{LPCODE} \citep{2005A&A...435..631A,2016A&A...588A..25M}. The version of \texttt{LPCODE} used in this work has a nuclear network that comprises 31 species and 96 reactions. We then post-process these models with a detailed nucleosynthesis multi-zone post-processing code, \texttt{ANT}: Astrophysical Nucleosynthesis Tool, using a reaction network of 1190 nuclear species, including isotopes with a half-life time down to $0.5\,$ seconds, in order to be able to follow the nucleosynthesis when neutrons densities are as high as $N_n=10^{17}\,$cm$^{-3}$. \texttt{ANT} has been developed during the realization of the present work.  In what follows we discuss the main features of both codes.

\subsection{Stellar evolution code and models}

The 31 species followed in \texttt{LPCODE} are: $n$, $^1$H, $^2$H, $^3$He, $^4$He, $^7$Li, $^7$Be, $^{12}$C , $^{13}$C, $^{14}$C, $^{13}$N, $^{14}$N, $^{15}$N, $^{16}$O, $^{17}$O, $^{18}$O, $^{19}$F, $^{20}$Ne, $^{21}$Ne, $^{22}$Ne, $^{23}$Na, $^{24}$Mg, $^{25}$Mg, $^{26}$Mg, $^{26}$Al, $^{26{\rm m}}$Al, $^{27}$Al, $^{28}$Si, $^{29}$Si, $^{30}$Si and $^{31}$P, where  $^{26{\rm m}}$Al is an excited meta-stable state of $^{26}$Al and $n$ are neutrons. The 96 reactions included in the code follow H and He burning including the hot CNO cycle, and the most relevant $\alpha$ and neutron-capture reactions. All the nuclear reaction rates are taken from NACRE II \citep{2013NuPhA.918...61X}, while the screening corrections are calculated following \cite{1973ApJ...181..439D}, \cite{1973ApJ...181..457G} and \cite{1982ApJ...258..696W}. Convective mixing is treated as a diffusive process following the Mixing Length Theory (MLT, \citealt{1932ZA......5..117B,1958ZA.....46..108B}) for the calculation of the diffusion coefficient. We do not take into account convective boundary mixing (CBM) in this work. The abundance changes due to nuclear burning and mixing are calculated simultaneously in a coupled way. 

We have calculated the evolution of stellar models from the zero-age main sequence to the red-giant branch (RGB). All of our models have an initial mass of $1\,M_{\odot}$, and three different values of initial composition (see Table \ref{tab:runs}). In order to mimic a mass-loss stage, we have enhanced the mass loss close to the tip of the RGB, up to a value of $10^{-5}\,M_{\odot}/{\rm yr}$. 
This is slow enough so that the structure of the envelope remains close to thermal equilibrium (i.e. steady state) and the location of the star in the HR diagram is univocally determined by the remaining envelope mass. 
The behaviour of the models after the stripping episode depends on the relation between the core temperature and the H-shell burning energy release for a given total mass left after the stripping. 
It should be noted that if the mass transfer happens at a much faster pace, envelopes would depart from a steady state and results can differ significantly from those presented in this work \citep[see for example the work of][]{2013MNRAS.435.2048H}. 
Also, our models can represent the evolution of stripped RGB stars in binaries as long as the posterior evolution of the core is not affected by the companion in a very close binary system.

\subsection{Nucleosynthesis code and calculations}
\begin{figure}
	% To include a figure from a file named example.*
	% Allowable file formats are eps or ps if compiling using latex
	% or pdf, png, jpg if compiling using pdflatex
	\includegraphics[width=1\columnwidth]{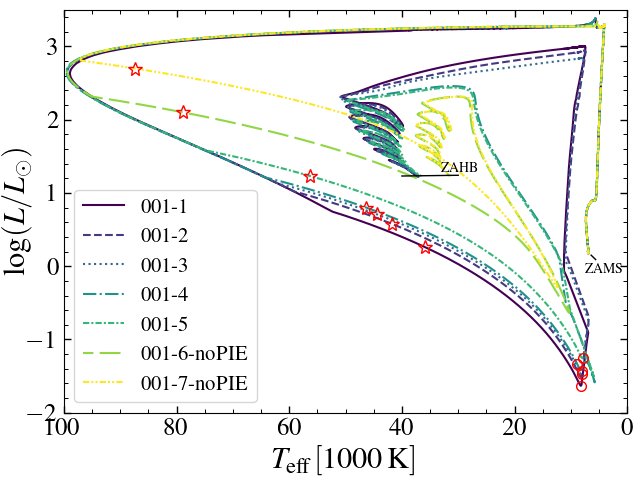}
    \caption{Hertzsprung-Russel diagram of the evolutionary sequences with initial metallicity $z = 0.001$ from the zero-age main sequence (ZAMS) to the zero-age horizontal branch (ZAHB). See Table \ref{tab:runs} for details about the different evolutionary sequences. The red stars mark the location of the peak of the He-flash for each sequence. The red circles mark the maximum energy liberation of CNO burning in those sequences that experience a PIE.}
    \label{fig:hr}
\end{figure}

\texttt{ANT} solves the nuclear reaction equations using the Bader-Deufhard method \citep{bader1983semi}, which is essentially a Bulirsch-Stoer method modified for stiff problems. The linear algebra package for solving the system is the MA28 package \citep{reid1986direct} that uses a direct method for sparse matrices. This implementation follows the suggestion of \cite{1999ApJS..124..241T}. The reaction rates for weak interactions (electronic captures and $\beta$ decays) are mainly taken from \cite{1987ADNDT..36..375T}, while all the other reaction rates are taken from the database of JINA Reaclib \citep{2010ApJS..189..240C}. For the neutron-capture reactions, the suggestion of JINA Reaclib at the moment of the realization of this work is a compromise between the KadOnis v.03 reaction rates when available \citep{dillmann2009kadonis,2006AIPC..819..123D} and theoretical reactions rates otherwise, mainly taken from \cite{2000ADNDT..75....1R,2010ApJS..189..240C,2010A&A...513A..61P}. In \texttt{ANT} we include the screening corrections for all the reactions involving two particles as reactants, following the works of \cite{1973ApJ...181..439D} and \cite{1973ApJ...181..457G}. We also include an electron screening correction for the $3\alpha$ reaction. We calculate this correction following the suggestion of \cite{1969ApJ...155..183S}. The nuclear network of \texttt{ANT} can comprise up to 5000 species. As stated above, for this work we have used a reaction network of 1190 nuclear species up to Po ($Z = 84$), including isotopes with a half-life time down to $0.5\,$ seconds. 

Convective mixing in \texttt{ANT} follows the mixing scheme of \cite{2001ApJ...554.1159C} (see also \citealt{2006NuPhA.777..311S,2009ApJ...696..797C}). In this scheme, the abundances per mass fraction of an isotope $k$ in a shell $i$ after the mixing in a time step $\Delta t$, $X^k_i$, are calculated from the abundances before the mixing $^0X^k_j$ of the same isotope $k$ in all the convective shells $j$ as:
\begin{equation}
    X_i^k =\,^0\!X_i^k + \frac{1}{M_{\rm conv}} \sum_{j = {\rm conv}} (^0\!X_j^k-\,^0\!X_i^k)f_{ij} \Delta m_j,\label{eq:chieffi}
\end{equation}
where $\Delta m_j$ is the mass of the shell $j$, $M_{\rm conv}$ is the total mass of the convective region, and $f_{ij}$ is a ``damping factor'' defined as:
\begin{equation}
    f_{ij} = \min\left(\frac{\Delta t}{\tau_{ij}},1\right),
\end{equation}
where $\Delta t$ is the time step in which the material is mixed and $\tau_{ij}$ is a characteristic time scale of convective motions between shells $i$ and $j$ that we take as:
\begin{equation}
    \tau_{ij} = \int_{r_i}^{r_j} \frac{dr}{v_{\rm MLT}}  = \sum_{l=i}^j \frac{\Delta r_l}{v_{l,{\rm MLT}}},
\end{equation}
being  $v_{l,{\rm MLT}}$ the convective velocity of shell $l$, which are taken from the stellar evolution calculations performed with \texttt{LPCODE}. 

To perform the post-processing computations with \texttt{ANT} we read the composition of the first model to be post-processed from \texttt{LPCODE}. For this initial model, the abundances of the 31 species followed by \texttt{LPCODE} are directly adopted in \texttt{ANT}, and the rest of the species are taken to be solar scaled or solar scaled with an $\alpha$ enhancement, depending on the specific parameters of the stellar model. From that point onwards, the chemical evolution is computed with the full \texttt{ANT} network and adopting the thermal and density structure together with the mixing velocities from the \texttt{LPCODE} models. Burning and mixing are treated in a decoupled way in \texttt{ANT}. Between two evolutionary time steps in \texttt{LPCODE}, \texttt{ANT} performs a number of sub-timesteps alternating the computation of the nuclear burning and the mixing of material. This approach should converge to a coupled treatment when the sub-timesteps converge to zero. 
In this work, we perform  with \texttt{ANT} five sub-timesteps between two \texttt{LPCODE} models. The time step of \texttt{LPCODE} in the proton-ingestion episode is of the order of 15 minutes, and therefore, \texttt{ANT} time steps are of the order of 3 minutes. The convective turn-over timescale calculated as the integral of $1/v_{\rm MLT}$ over the whole convective zone at the beginning of the PIE varies in the models from $\sim2\,$h to $\sim 24\,$h. 
In order to assess the error in the abundances due to this approach, we have recalculated the chemical evolution of one sequence with 100 timesteps in \texttt{ANT} for each evolutionary timestep in \texttt{LPCODE} (sequence 001-4, see Table \ref{tab:runs}). The difference on the surface abundances obtained are of the order of $0.04\,$dex, having the majority of the elements a difference below $0.1\,$dex. The exceptions are Be, Rb, Xe and Cs. The greatest difference was encountered for Xe, of about $0.33\,$dex.

The detailed chemical evolution was followed with \texttt{ANT} up to the point where the nucleosynthesized material is dredged up to the surface of the star. Afterward, abundances of unstable isotopes were decayed for about $1.5-2\,$Myr to obtain the abundances at the zero-age horizontal branch.

\begin{table*}
\caption{Properties of the evolutionary sequences computed in this work.}
\begin{tabular}{l c c c c c c c c c c} 
\hline
Name &$\!\!\!\!$z$_{\rm ini}$ $\!\!\!\!$&$\!\!\!\!$ y$_{\rm ini}$  $\!\!\!\!$&$\!\!\!\!$ PIE? $\!\!\!\!$&$\!\!\!\!$  $M_{\star}/M_{\odot}$ $\!\!\!\!$&$\!\!\!\!$ $\log\left(\frac{M_{\rm H}}{M_{\odot}}\right)_{\rm L_{He}=1}$ $\!\!\!\!$& $\!\!\!\!$ $\log\left(\frac{L_{\rm CNO}}{L_{\odot}}\right)_{\rm L_{He}=1}$ $\!\!\!\!$& $\!\!\!\!$ $\Delta S_{\rm{max\,of\,}L_{\rm He}}^{[N_{\!\rm A}k_{\rm B}/\mu]}$ $\!\!\!\!$&       $\!\!\!\!$ $t_{\rm split} - t_{\rm PIE}$ $\!\!\!\!$&$\!\!\!\!$ $N_{n,{\rm max}}\,[{\rm cm}^{-3}]$ $\!\!\!\!$&$\!\!\!\!$ $\tau_{n,\rm max}\,$[mb$^{-1}$]\\
\hline
01-1&$0.01$ & $0.265$ & yes & $0.4667$& $-3.4728$& $-1.442$& $1.31$& $\sim 26\,$d & $1.2\times 10^{10}$ & $1.9\times 10^{-4}$\\
01-2&$0.01$ & $0.265$ & yes & $0.4720$& $-3.4887$& $-0.849$& $1.63$ & $\sim 110\,$d & $4.6\times 10^{9}$ & $1.5\times 10^{-4}$\\
01-3&$0.01$ & $0.265$ & yes & $0.4745$& $-3.4158$& $3.016$& $2.24$ & $\sim 193\,$yr & $4.2\times 10^{9}$ & $9.9\times 10^{-4}$\\
01-4&$0.01$ & $0.265$ & yes & $0.4747$& $-3.3760$& $3.103$& $2.32$ & $\sim 620\,$yr & $7.4\times 10^{8}$ & $8.9\times 10^{-4}$\\
01-5&$0.01$ & $0.265$ & no & $0.4748$& $-3.3128$& $3.191$& $2.55$ & - & - & -\\
01-6&$0.01$ & $0.265$ & no & $0.4749$& $-3.2951$& $3.211$& $2.75$& - & - & -\\
01-7&$0.01$ & $0.265$ & no & $0.4751$& $-3.1960$& $3.287$& $4.14$ & - & - & -\\
\hline
001-1&$0.001$ & $0.246$ & yes & $0.4821$& $-3.2958$& $-0.443$& $0.91$& $\sim 28\,$d & $1.6\times 10^{12}$ & $7.4\times10^{-3}$\\
001-2&$0.001$ & $0.246$ & yes & $0.4835$& $-3.2956$& $1.511$& $1.09$ & $\sim 172\,$d & $4.9\times 10^{11}$ & $6.1\times 10^{-3}$\\
001-3&$0.001$ & $0.246$ & yes & $0.4836$& $-3.2772$& $2.674$& $1.16$ & $\sim 1\,$yr & $8.0\times 10^{11}$ & $9.8\times 10^{-3}$\\
001-4&$0.001$ & $0.246$ & yes & $0.4838$& $-3.2593$& $2.794$& $1.21$ & $\sim 400\,$yr& $3.0\times 10^{13}$ & $2.21$\\
001-5&$0.001$ & $0.246$ & yes & $0.4839$& $-3.2220$& $2.919$& $1.46$ & $\sim 670\,$yr & $1.7\times 10^{12}$ & $5.12$\\
001-6&$0.001$ & $0.246$ & no & $0.4840$ & $-3.1903$ & $2.986$ & $1.97$ & - & - & - \\
001-7&$0.001$ & $0.246$ & no & $0.4843$ & $-3.1215$ & $3.081$ & $2.44$ & - & - & - \\
\hline
0001-1&$0.0001$ & $0.2451$ & yes & $0.4900$& $-3.1305$& $-0.790$& $1.03$ & $\sim 0.9\,$d, $17\,$d & $1.2\times 10^{15}$ & $2.98$\\
0001-2&$0.0001$ & $0.2451$ & yes & $0.4912$& $-3.1338$& $-0.682$& $1.02$ & $\sim 1.6\,$d & $5.5\times 10^{14}$ & $3.57$\\
0001-3&$0.0001$ & $0.2451$ & yes & $0.4926$& $-3.1345$& $-0.481$& $1.23$ & $\sim 1.6\,$d & $1.0\times 10^{15}$ & $3.74$\\
0001-4&$0.0001$ & $0.2451$ & yes & $0.4939$& $-3.1411$& $-0.227$& $1.46$ & $\sim 2.7\,$d & $8.2\times 10^{14}$ & $2.40$\\
0001-5&$0.0001$ & $0.2451$ & yes & $0.4946$& $-3.1421$& $0.083$& $1.48$ & $\sim 4.6\,$d & $8.6\times 10^{14}$ & $1.61$\\
0001-6&$0.0001$ & $0.2451$ & yes & $0.4948$& $-3.1422$& $0.271$& $1.53$ & $\sim 6\,$d & $7.7\times 10^{14}$ & $1.59$\\
0001-7&$0.0001$ & $0.2451$ & yes & $0.4951$& $-3.1306$& $0.702$& $1.60$ & $\sim 9\,$d & $6.4\times 10^{14}$ & $1.76$\\
0001-8&$0.0001$ & $0.2451$ & yes & $0.4953$& $-3.1395$& $2.402$& $1.73$ & $\sim 17\,$d & $4.9\times 10^{14}$ & $2.09$\\
0001-9&$0.0001$ & $0.2451$ & yes & $0.4956$& $-3.1159$& $2.630$& $2.05$ & $\sim 117\,$d & $2.2\times 10^{14}$ & $2.86$\\
0001-10&$0.0001$ & $0.2451$ & yes & $0.4957$& $-3.1027$& $2.682$& $2.57$ & $\sim 264\,$yr & $1.6\times 10^{13}$ & $8.41$\\
0001-11&$0.0001$ & $0.2451$ & no & $0.4958$ & $-3.0862$ & $3.25$ & $2.744$ & - & - & - \\
0001-12&$0.0001$ & $0.2451$ & no & $0.4965$ & $-2.9714$ & $3.62$ & $2.974$ & - & - & - \\

\hline
\end{tabular}
\tablefoot{In order the columns are: name of the evolutionary sequence, initial metallicity, initial He content, if they experience or not a proton ingestion episode, remnant mass, hydrogen mass and CNO burning luminosity at the moment when the He-burning luminosity reaches the value $\log(L_{\rm He}/L_{\odot})=1$, specific entropy barrier at the peak of the He flash, time span between the onset of the PIE and the splitting of the convective zone, maximum neutron density and maximum neutron exposure reached during the evolution.}
\label{tab:runs}
\end{table*}

%______________________________________________________________

\section{Evolution of late core-He flashes}\label{sec:vlhf}

\begin{figure*}
	% To include a figure from a file named example.*
	% Allowable file formats are eps or ps if compiling using latex
	% or pdf, png, jpg if compiling using pdflatex
 \centering
	\includegraphics[width=2\columnwidth]{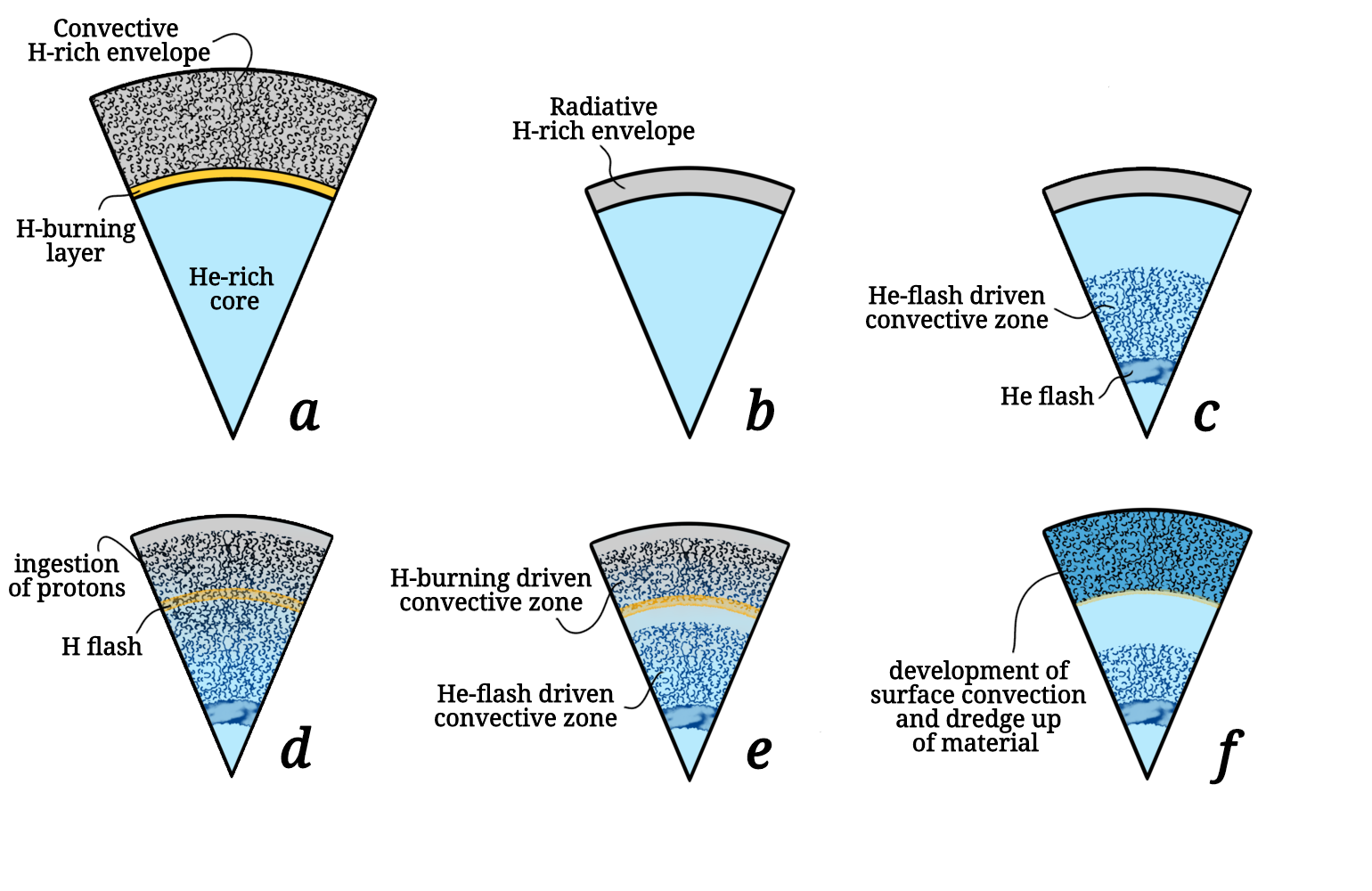}
    \caption{Schematic description of the main stages of the development of a proton ingestion episode in a He-core flash model. Grey and light-blue regions correspond to H-rich and He-rich regions, respectively.  Yellow indicates  H-burning regions and clouded regions indicate the presence of convection.}
    \label{fig:pizza}
    
\end{figure*}
\begin{figure}

	% To include a figure from a file named example.*
	% Allowable file formats are eps or ps if compiling using latex
	% or pdf, png, jpg if compiling using pdflatex
	\includegraphics[width=\columnwidth]{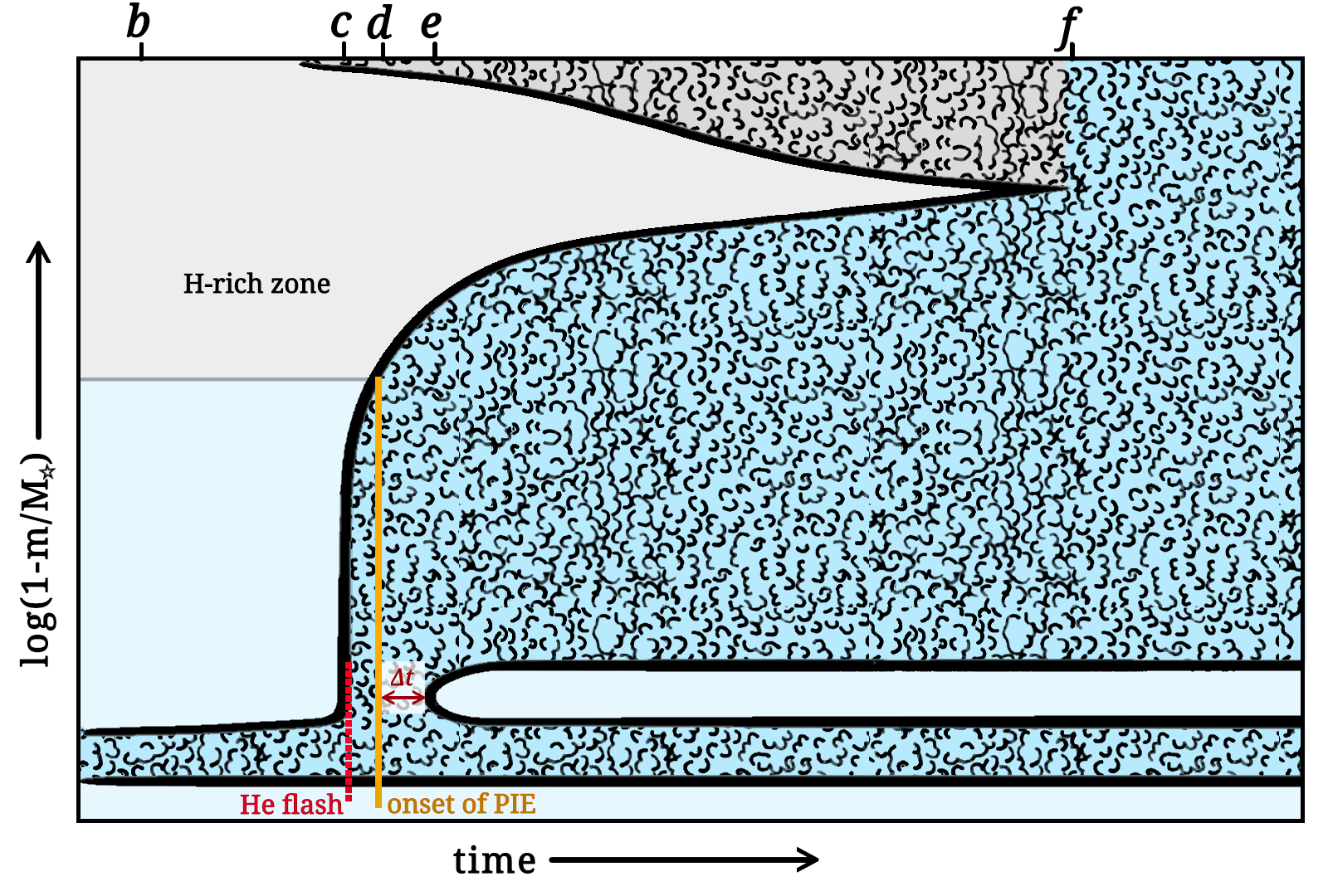}
    \caption{Schematic Kippenhahn diagram of the evolution of a proton ingestion episode in a He-core flash model.}
    \label{fig:kipp:sketch}
    
\end{figure}

In the canonical single-stellar evolution picture, the He-flash takes place when the star reaches the tip of the red-giant branch. However, under certain conditions, a star can be stripped of most of its envelope before it reaches the core temperature needed for He ignition. This can in principle happen in both single stars, if they have low initial metallicity and high He content, and in interactive binaries due to stable mass transfer or mass loss in common envelope episodes. When this happens, the star departs from the RGB contracting towards higher effective temperatures at constant luminosity (see Fig.~\ref{fig:hr}). 
The luminosity of the star is sustained by the burning of H at the bottom of the remaining envelope. As the envelope is consumed, the core keeps increasing its mass and its temperature rises accordingly. 
 \cite{1993ApJ...407..649C} demonstrated that a He flash can still develop when the star is entering the WD cooling curve. Depending on the mass of the remaining H-rich envelope, stars that undergo a late He core flash will populate the blue ($T_{\rm eff}>7000\,$K) or extreme ($T_{\rm eff}> 22000\,$K) horizontal branches during their quiescent He burning phase. 
The majority of hot-subdwarf stars are located on the extreme horizontal branch \citep{2016PASP..128h2001H}, and many works have employed late core-flashes models to study these stars (e.g. \citealt{1996ApJ...466..359D,2001ApJ...562..368B, 2003ApJ...582L..43C, 2004ApJ...602..342L, 2006A&A...457..569C, 2008A&A...491..253M, 2017A&A...599A..54X, 2018A&A...614A.136B, 2018MNRAS.475.4728B, 2018MNRAS.481.3810B}). Observations of hot-subdwarf stars are qualitatively very well described by these models (e.g. \citealt{2008A&A...491..253M, 2017A&A...599A..54X}), suggesting that many of these compact objects have lost most of their envelope mass near the tip of the red giant branch, probably by binary interactions \citep{2020A&A...642A.180P}. He-rich hot subdwarfs, in particular, can be explained by a late core He-flash that happens when the H-burning shell is almost extinct, 
allowing the He-flash driven convective zone to reach the H-rich envelope. The remaining H is then engulfed and violently burnt in the interior of the star \citep{2008A&A...491..253M}. 
Another possible scenario for the formation of He-rich hot subdwarfs is the merger of two He-core WDs or a He-core WD and a low-mass CO core WD (see \citealt{2011MNRAS.410..984J, 2012MNRAS.419..452Z, 2018A&A...620A..36S, 2022MNRAS.511L..60M} for discussions). In this work, we focus on models of late core-He flashes that can lead to proton-ingestion episodes. However, it is worth noting that in the merger scenario, the post-merger evolution can also include proton ingestion episodes \citep{2002MNRAS.333..121S}, and therefore the possibility of $i$-process nucleosynthesis as studied in this work.

\subsection{Qualitative evolution of proton-ingestion episodes in very late hot-flashers}
\label{sec:qualitative-evol-PIE}

The evolution of He flashes in stripped post-RGB stars can be qualitatively separated into three different types. 
These types have been termed in previous works as early hot flasher, hot flasher with shallow mixing and  hot flasher with deep mixing, which mainly depend on how much mass remains in the H-rich envelope and how active the H-burning shell is when the He flash develops (\citealt{2001ApJ...562..368B, 2004ApJ...602..342L}, see \citealt{2008A&A...491..253M, 2018A&A...614A.136B} for further details about these different behaviors).

If the H-rich layer is active enough, the material around it is stable against convective motions and the convective region developed by the He-flash does not penetrate the H-rich layers. This behavior is often described in terms of an `entropy barrier' for convection due to the existence of the H-burning shell \citep{1976ApJ...208..165I}. These are the so-called early hot flasher and shallow mixing cases. 
We will discuss the entropy barrier in our models in Section \ref{sec:PIEornotPIE}. After the He-flash, and due to the energy liberated by it, the low-mass envelope expands and the surface of the star becomes cooler and dimmer  (see Fig. \ref{fig:hr}). As the temperature of the shell is inversely proportional to its radius \citep[see Section \S 33.2 in ][]{2012sse..book.....K}, 
the expansion leads to a decrease in the temperature of the burning shell and the surface luminosity drops. The expansion also leads to the formation of an outer convective envelope as the effective temperature of the star drops down. In early hot flashers, the surface composition of the star remains completely unchanged, as the convective envelope never reaches the material processed by the H-burning shell. However, in the shallow mixing case, the H-burning shell becomes extinct during this expansion phase and the convective envelope can reach down the H-deficient core, mixing some material processed via the CNO cycle to the surface. A significantly different behavior occurs if the He flash develops when the H-burning shell is already almost extinct. These are the so-called deep mixing hot flashers, in which the He-flash driven convective zone is able to reach the H-rich layers, leading to a proton ingestion event (PIE). 

In Figs \ref{fig:pizza} and \ref{fig:kipp:sketch} we show, schematically, the main stages of the evolution of the star before and after a PIE. Each panel of Fig. \ref{fig:pizza} corresponds to a key stage in the development of the PIE, and is marked in the schematic Kippenhahn diagram displayed in Fig. \ref{fig:kipp:sketch}. If the RGB star (Fig.~\ref{fig:pizza}$a$) loses most of its H-rich envelope, down to or less than $\sim 5\times10^{-4}\,M_{\odot}$ (depending on the initial metallicity) the H-burning layer dims before the He-flash occurs (Fig.~\ref{fig:pizza}$b$). Because of the dimming of the H-burning layer the luminosity of the star decreases and enters the white dwarf cooling phase. 
If the temperature of the core reaches the critical value for igniting He at the start of the WD cooling phase 
(Fig.~\ref{fig:pizza}$c$), the convective region driven by the He-flash can penetrate 
%aprende a escribir, MArcelo
into the H-rich layers. Protons are thus ingested into the hot, C- and He-rich interior, where they are rapidly burned producing a H flash (Fig.~\ref{fig:pizza}$d$). 
As energy is released by H burning, the integrated heat flux $l(r)$ drops in the convective layers immediately below, causing the radiative temperature gradient $\nabla_{\rm rad}$ to drop. This leads to a drop in the convective velocities required to transport the excess heat not transported by radiation. In these inner convective layers the temperature gradient is almost the adiabatic one, $\nabla_{\rm ad}$. 
Finally, when $l(r)$ drops enough so that $\nabla_{\rm rad}=\nabla_{\rm ad}$, convective motions stop according to the classical Schwarzschild criterion \citep{1906WisGo.195...41S}. 
This leads to the splitting of the He-flash driven convective zone into two (Fig.~\ref{fig:pizza}$e$). When this happens, no more fresh H is ingested into the deep He burning region. At this point, the structure of the models usually has three convective zones, an inner one driven by the still active He-flash, an intermediate one driven by the H-burning, and an external convective zone developed at the surface, which deepens as the photosphere cools down. Once the peak of the H-flash occurs, the surface luminosity of the star begins to rise again while the effective temperature continues to decrease until the external convective zone reaches the intermediate convective zone (Fig.~\ref{fig:pizza}$f$). The intermediate convective zone has material that has been processed in both the He and H flashes, and therefore the surface of the star is enriched with nucleosynthesized material.

\subsubsection{Nucleosynthesis and neutron production during a PIE}

\begin{figure}
	% To include a figure from a file named example.*
	% Allowable file formats are eps or ps if compiling using latex
	% or pdf, png, jpg if compiling using pdflatex
	\includegraphics[width=1\columnwidth]{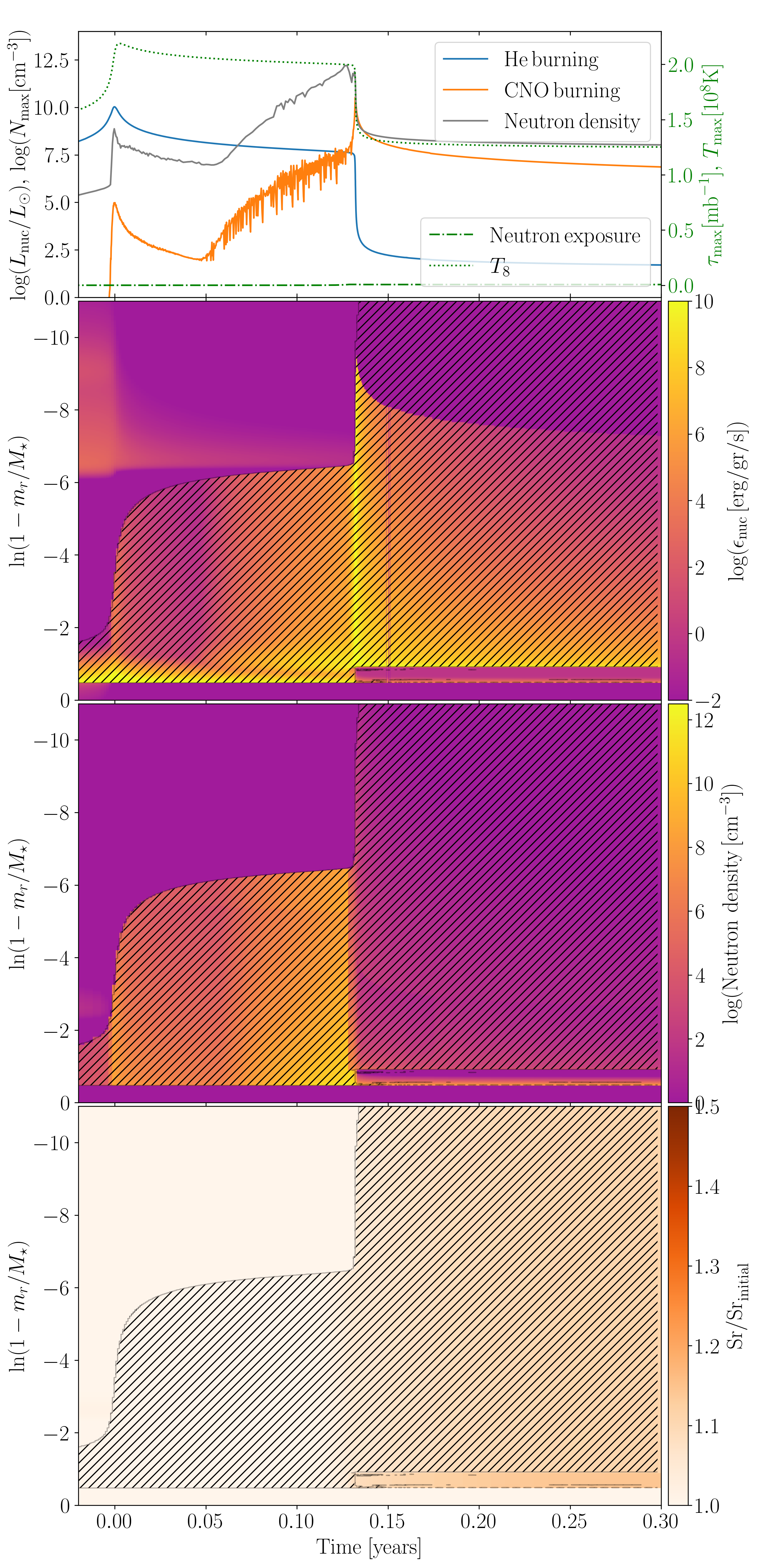}
    \caption{Kippenhahn diagrams for model 001-1. The upper panel shows the evolution of the He burning luminosity (blue line), the CNO burning luminosity (orange line), and the neutron density (gray line). The right y-axis corresponds to the neutron exposure, shown in a dot-dashed green line, and the maximum temperature of the model at each time, shown in a dotted green line. In the Kippenhahn diagrams (the three lower panels) the convective zones are indicated as black dashed areas. The uppermost Kippenhahn diagram shows in color code the nuclear energy liberation. The middle Kippenhahn diagram shows the neutron density, and the lowermost Kippenhahn diagram shows the abundance of Sr respect to its initial value.}
    \label{fig:KD:001-1}
\end{figure}

\begin{figure}
	% To include a figure from a file named example.*
	% Allowable file formats are eps or ps if compiling using latex
	% or pdf, png, jpg if compiling using pdflatex
	\includegraphics[width=1\columnwidth]{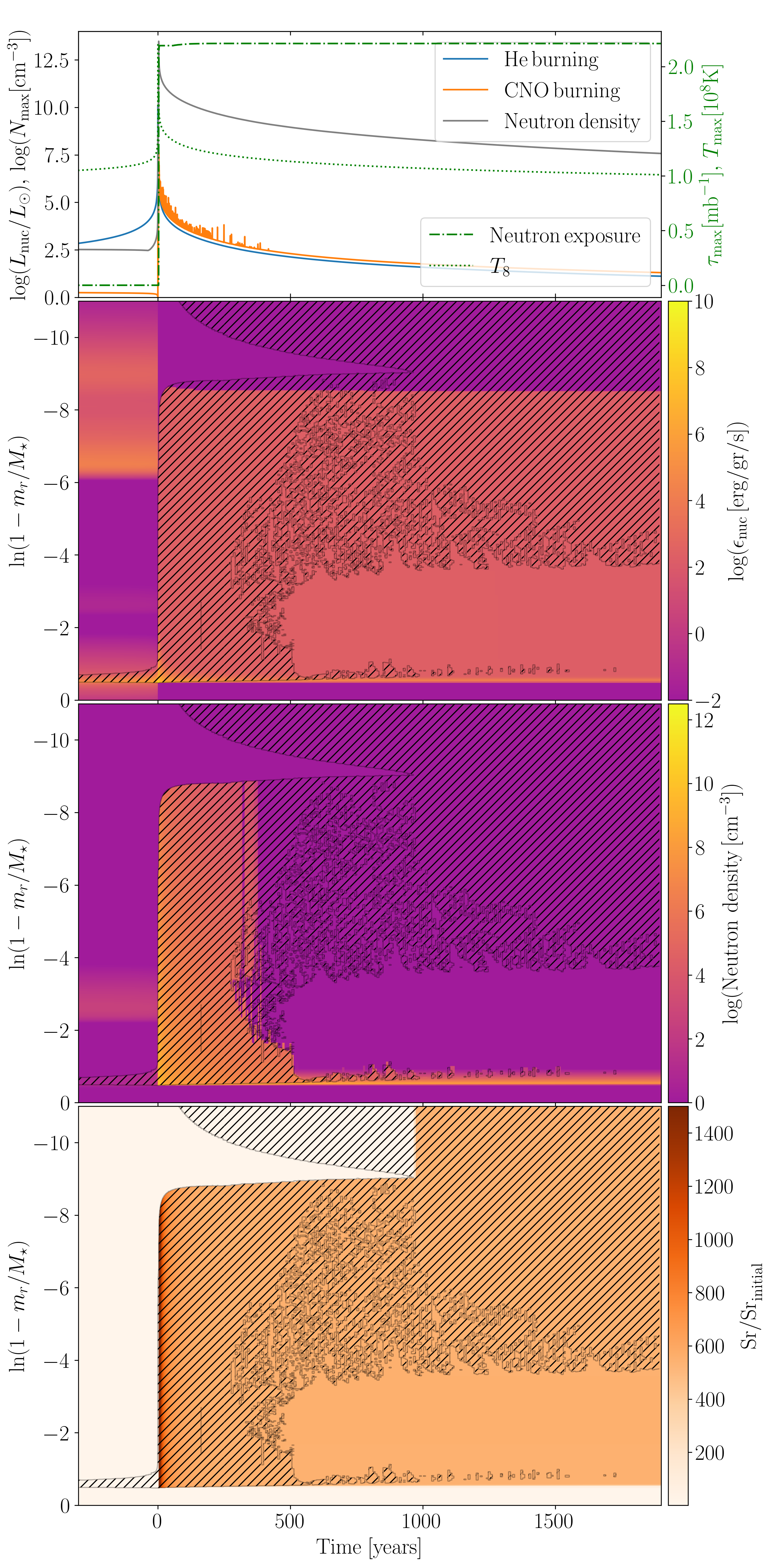}
    \caption{Same as Fig. \ref{fig:KD:001-1} but for model 00-1-4.}
    \label{fig:KD:001-4}
\end{figure}

\begin{figure}
	% To include a figure from a file named example.*
	% Allowable file formats are eps or ps if compiling using latex
	% or pdf, png, jpg if compiling using pdflatex
	\includegraphics[width=1\columnwidth]{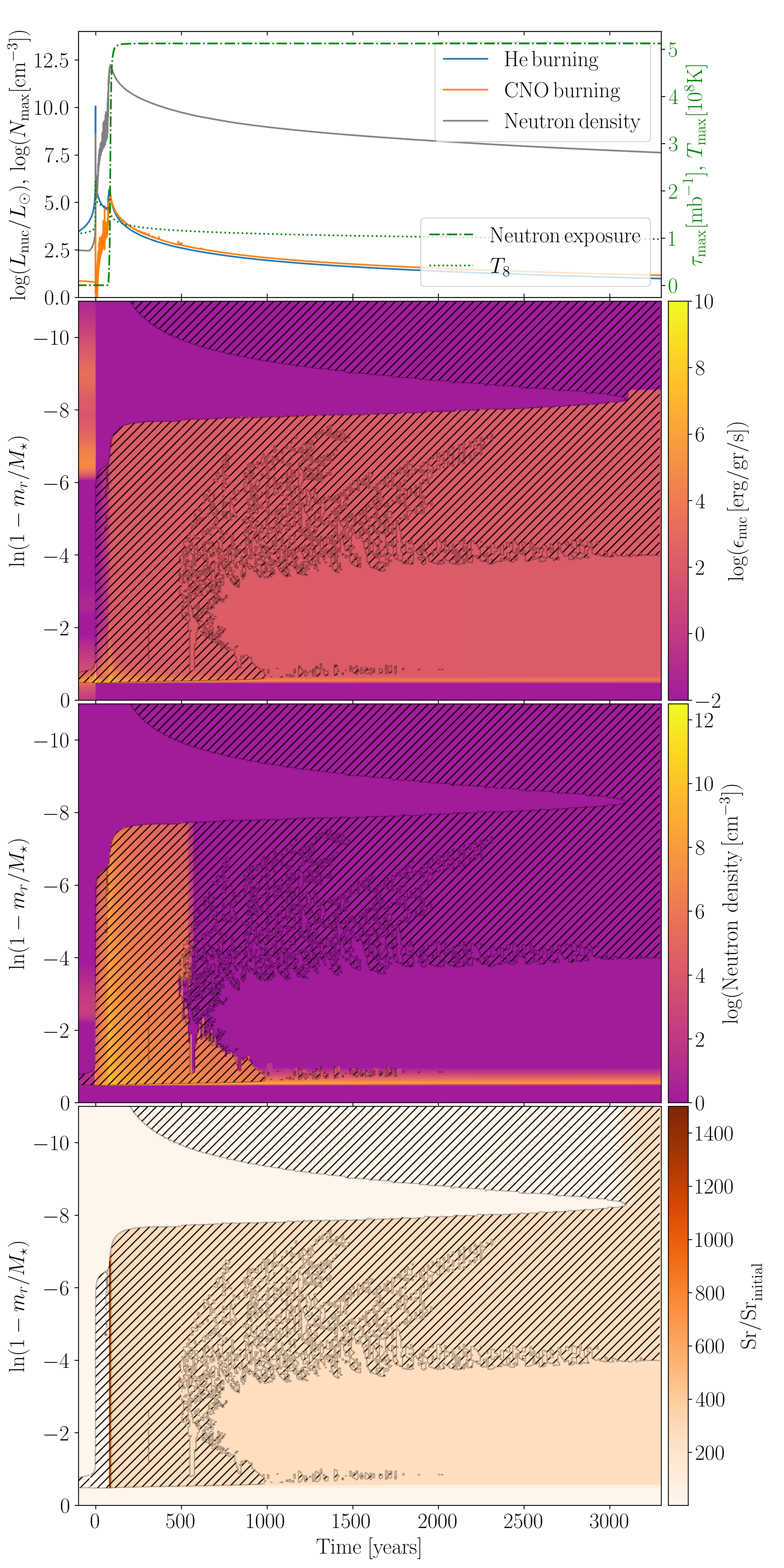}
    \caption{Same as Fig. \ref{fig:KD:001-1} but for model 00-1-5.}
    \label{fig:KD:001-5}
\end{figure}

In proton-ingestion episodes, the protons are ingested into the hot interior where He burning is taking place, an environment rich in He and also $^{12}$C. At the hot temperatures of this region, protons ($p$) quickly 
react with $^{12}$C, leading to the production of $^{13}$N  and, through $\beta^-$-decay, of $^{13}$C. At the high temperatures of He burning $^{13}$C is destroyed by $\alpha$ captures, releasing a neutron ($n$) in the process.
The chain reaction leading to the production of neutrons is, thus
\begin{equation}
    ^{12}{\rm C}(p,\gamma)^{13}{\rm N}\,(\beta^-)^{13}{\rm C}(\alpha,n)^{16}{\rm O}.
\end{equation}
Protons can also react with the recently created $^{13}$C leading to the formation of $^{14}$N.
The production of $^{14}$N not only removes protons from the stellar plasma, decreasing the production of $^{13}$C, but also allows the capture of neutrons by $^{14}$N, reducing the net production of free neutrons. As such the formation of $^{14}$N is a very efficient poison for free neutrons. 
All of this nucleosynthesis happens in the convective zone driven by the He-flash, mainly in the hot bottom. When the convective zone splits into two, no more protons are being ingested into the hot region where the temperature is high enough for $^{13}$C to react with He. 
Therefore, the nucleosynthesis of heavy elements can only continue until the already existing $^{13}$C in this region is exhausted.  The neutron density in this phase can be order of magnitudes lower than before the splitting of the convective zones (see Figs. \ref{fig:KD:001-1}, \ref{fig:KD:001-4}, and \ref{fig:KD:001-5}). 
The elements synthesized in this inner convective zone after the splitting will not appear at the surface of the star\footnote{Unless mixing happens across the thin stable region that develops in 1D stellar models. Notably, the absence of this stable region has been found in 3D hydrodynamical simulations of the earlier stages of the PIE \citep{2012ASSP...26...87M}.}. Protons continue to be ingested into the intermediate convective zone, but there the temperature soon drops below He burning, and therefore, $^{13}$C cannot react anymore with He. 
Neutron production in this region is consequently immediately halted when the convective zones splits. 
This behavior can be appreciated in Figs. \ref{fig:KD:001-1}, \ref{fig:KD:001-4}, and \ref{fig:KD:001-5}, where we show the evolution of the neutron density in Kippenhahn diagrams. The intermediate convective zone, already enriched in heavy elements will later be reached by the outer convective zone and the material will be mixed to the surface. Therefore, the timing of the splitting of the convective zones is key for the final pollution of the stellar photosphere with heavy elements. The outer convective zone does not harbour a significant mass 
%doe not represent a significant fraction of the total mass, 
and therefore the processed material is not much diluted  (see Figs. \ref{fig:KD:001-1}, \ref{fig:KD:001-4}, and \ref{fig:KD:001-5}). The surface of the hot-subdwarf star becomes enriched with the material processed in the interior, meaning it will present enhancement in He, C, N, Ne, and (in some cases) heavy elements. \cite{2008A&A...491..253M} and \cite{2018A&A...614A.136B} have calculated in detail the surface enrichment in light elements after PIEs up to O and Ne respectively. We will focus in this work on the synthesis of trans-irons elements through neutron captures.

\subsection{Description of our models}

The metallicity of hot-subdwarf stars cannot be directly measured because diffusion processes take place in their atmospheres, washing away the original composition of the material. However, \cite{2021A&A...653A.120D} have identified a heavy-metal hot subdwarf, EC 22536-5304, in a binary system with a subdwarf F-type (sdF) star. They reported for the sdF star a metallicity of $[{\rm Fe}/{\rm H}] = -1.95\pm 0.04$, with an $\alpha$-enhanced composition of $[{\rm \alpha}/{\rm Fe}]=0.4\pm 0.04$. The chemical composition adopted in \cite{2023A&A...680L..13B} was motivated by this finding. Another heavy-metal hot subdwarf having a binary companion is SB 744 \citep{2021A&A...653A...3N}, with an observed metallicity of $[{\rm Fe}/{\rm H}] = -1.02\pm 0.15$. However, this star has a significantly low helium abundance, which can suggest that diffusion processes have been acting on its surface for an extended period. In this work, we want to assess the differences on the neutron capture nucleosynthesis due to metallicity, and the mass of the H-rich envelope. We have performed a set of simulations with initial metallicities $Z=0.01$, $0.001$ and $0.0001$ and solar-scaled metal abundances. 
Each set of runs covers a range of envelope masses at the moment of the He flash that allows us to sample all possible types of the late hot-flashers, focusing on those who experience PIEs. Even if not presented here in detail, our models cover also the early hot-flashers and the formation of He-core WDs, and can be provided upon request. 

In Table \ref{tab:runs} we show key properties of our models, namely, their initial composition, whether they experience a PIE or not, total mass of the sequences after the mass-stripping episode, mass of H and CNO burning luminosity before the He-flash \footnote{More precisely when the He burning luminosity starts to increase and reaches the value $\log(L_{\rm He}/L_{\odot})=1$}, the value of the entropy barrier at the maximum energy release of the He flash, the time span between the onset of PIE and the splitting of the convective zone, the maximum neutron density and the maximum neutron exposure\footnote{The neutron exposure in our models is calculated averaging its value in the convective zone, see Appendix \ref{appx:1} for more details} reached during the evolution of the sequences. We can see from this table that within a small range of stellar masses, very different behaviors are possible. Models of similar masses may or may not experience a PIE, and when they do, the timescale between this episode and the splitting of the convective zone can differ by orders of magnitude. This fact directly affects the nucleosynthesis, changing the evolution of the neutron density and neutron exposure. In the next section we will discuss what determines the occurrence of a PIE and the splitting of the convective zone, and how this affects the final nucleosynthesis.

\subsubsection{PIE or not PIE}\label{sec:PIEornotPIE}

\begin{figure}
	% To include a figure from a file named example.*
	% Allowable file formats are eps or ps if compiling using latex
	% or pdf, png, jpg if compiling using pdflatex
	\includegraphics[width=1\columnwidth]{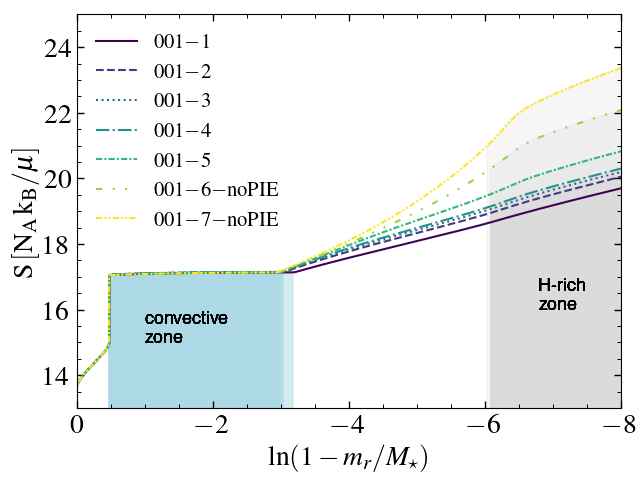}
 \includegraphics[width=1\columnwidth]{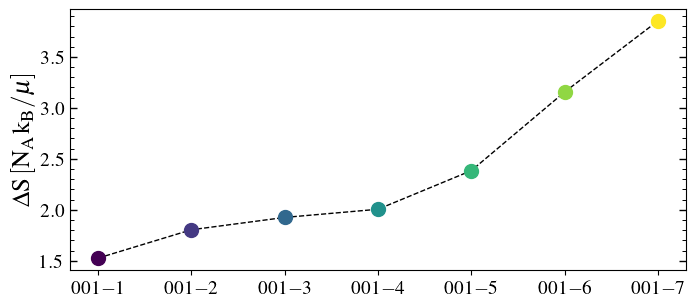}
    \caption{Upper panel: Specific entropy of all models with initial metallicity $z=0.001$ at the peak of the He flash. Lower panel: Entropy barrier at the peak of the He flash for the models shown in the upper panel.}
    \label{fig:entropy:all}
\end{figure}

\begin{figure}
	% To include a figure from a file named example.*
	% Allowable file formats are eps or ps if compiling using latex
	% or pdf, png, jpg if compiling using pdflatex
 	\includegraphics[width=1\columnwidth]{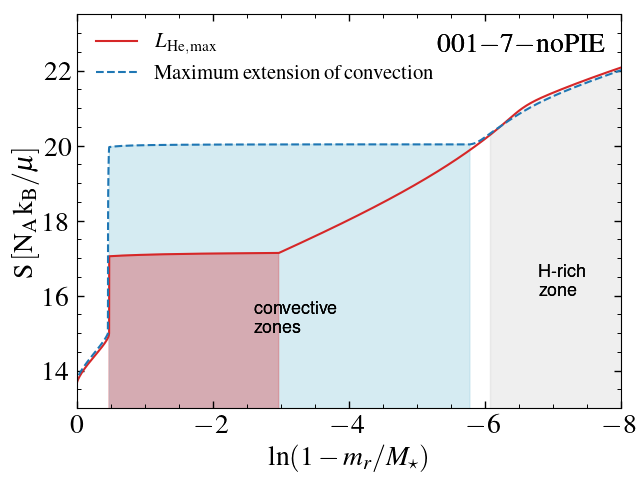}
    \includegraphics[width=1\columnwidth]{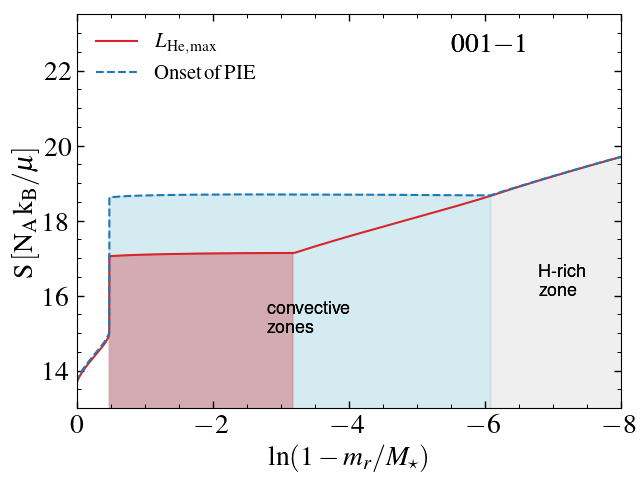}
    \caption{Lower panel: Specific entropy of model 001-1 at two different moments of the evolution. 
    The red line corresponds to the peak of the He flash, meanwhile the blue dashed line corresponds to the onset of the PIE. The red- and blue-shaded areas are the convective zones at each of these stages. The grey shaded area corresponds to the H-rich envelope. Upper panel: Specific entropy of model 001-7, which does not experience a PIE. Here the blue dashed line corresponds to the moment of the evolution where the convective zone reaches its maximum extension.}
    \label{fig:entropy:001-1-7}
\end{figure}

The existence of an `entropy barrier' below H-burning shells was shown by \cite{1976ApJ...208..165I} in the context of numerical models of thermally pulsating AGB stars. This `barrier' consists of an increase of the entropy at the base of the H-burning layer, which effectively prevents the thermal pulse-driven convective zone of AGB stars to reach the H-rich layers. The existence of a H-burning shell injects heat in that region of the star. As shown by \cite{1976ApJ...208..165I}, the structure of the star reacts to this heat until it reaches a steady state in which the entropy in the H-burning shell is higher than immediately below it. This can be understood by noting that in the regions immediately below a luminous burning shell\footnote{Here the term luminous indicates a burning shell that releases significantly  more power than the core below it, which is the standard situation in AGB and RGB stars.} an almost isothermal region of very little mass is formed. In this region, hydrostatic equilibrium forces the density to increase exponentially as we move inwards. For example, in the case of a classical monoatomic gas (radiation pressure always plays a minor role here) the density follows $\rho(r)\propto \exp{\mathcal{A} (R_s-r)/R_s\, r}$, where $\mathcal{A}$ is a positive constant with units of length and $R_s$ is the radius of the burning shell \citep[see][ for details]{2022ApJ...941..149M}. The specific entropy of a classical monoatomic ideal gas is $s=N_a k/\mu \ln{(T^{3/2}/\rho)}+\mathcal{C}$. Then, in a region of almost negligible mass the entropy increases  with radius as $s=\mathcal{B}(r-R_s)/R_s\, r+\mathcal{B}'$, where $\mathcal{B}$ and $\mathcal{B}'$ are two positive constants. As stated above, the existence of this almost massless region below the burning shell where the entropy increases sharply -the entropy `barrier'- has the consequence that the material there is effectively stable against convection. 
This can be understood by looking at the change of the specific entropy of the material written in terms of the local values of the temperature gradient, $\nabla$, and the adiabatic gradient, $\nabla_{\rm ad}$,  \cite{2004sipp.book.....H}),
\begin{equation}
    \frac{{\rm d}S}{{\rm d}r} = c_{\rm P}\,(\nabla-\nabla_{\rm ad})\,\frac{{\rm d}\ln P}{{\rm d}r}.
    \label{eq:dsdr}
\end{equation}
Eq. \ref{eq:dsdr} clarifies that in convective regions ($\nabla-\nabla_{\rm ad}\simeq 0$, according to the Schwarzschild's criteria) the entropy is almost a constant. Meanwhile, in stable regions  ($\nabla-\nabla_{\rm ad} < 0$), the entropy is a function that increases towards the exterior (if $c_P$ is positive, given that the pressure always decreases outwards). Eq. \ref{eq:dsdr} shows that the existence of a sharp rise in the value of $s$ creates a very stable region against convection, preventing the contact between the He-flash driven convective zone and the outer H-rich layers. As the existence of this barrier is a consequence of the presence of the H-burning shell, the contact between these layers becomes more and more probable the less luminous the burning shell. In models of thermal pulses of AGB stars, PIEs occur for models of low metallicity ($[$Fe/H$] < -2$, see \citealt{2022A&A...667A.155C}). In these cases, the low entropy value of the H-burning shell is due to the low metallicity. Lower CNO abundance in the H-burning shell leads to a lower rate of energy release by H-burning. In our models -and in general, in deep mixing models leading to the formation of hot-subdwarf stars-, not only the metallicity has an impact on the entropy barrier, but also the mass of the H-rich envelope. As the mass of the H-rich envelope becomes thinner, the H-burning layer becomes dimmer, making a PIE possible even at solar metallicities, as first shown by \cite{1977PASJ...29..331F}. 
Once the He-flash starts, the specific entropy of the material in the He-flash driven convective zone starts increasing, as the He-flash adds heat to the material in the convective zone. If the barrier is low enough (or the flash long/luminous enough) at some point the specific entropy of the material in the He-flash driven convective zone will reach that of the layers where H is still present. Convection then carries H-rich material down into the hot and C-enriched regions. In a specific He-core flash model, therefore, whether a PIE occurs or not is determined by the He-flash luminosity evolution and the luminosity of the H-burning shell immediately before the development of the He flash. 
The evolution of the He-flash is the same in all of our models of a given initial composition, as they are essentially identical in the interior. Consequently, the evolution of the entropy in the He-flash driven convective zone is also nearly identical across all models of the same metallicity (see Fig. \ref{fig:entropy:all}). To characterize the properties of each sequence we adopt the peak of the He-flash as a reference point, representing the same moment and entropy across all models. At this moment of the evolution, we calculate the entropy barrier as the difference between the entropy at the bottom of the H-rich layers and the convective zone: $\Delta S = S_{\rm conv} - S_{\rm H}$. Fig. \ref{fig:entropy:all} shows that the entropy barrier increases with higher H-envelope mass. Models 001-6 and 001-7 do not experience a PIE. For a PIE to occur, the specific entropy of the convective zone must reach the envelope's entropy during the He-flash evolution. If this does not happen and the entropy in the convective zone stops rising before reaching the envelope's value, a PIE does not occur. 
Fig. \ref{fig:entropy:001-1-7} illustrates the entropy for models 001-1 and 001-7 at two different moments. Model 001-1 has a thinner H envelope than 001-7; see Table \ref{tab:runs}  for further details. Model 001-1, which undergoes a PIE, is shown at the moment of maximum energy release during the He-flash and at the onset of the PIE. In contrast, model 001-7, which does not experience a PIE, is depicted at the peak He-flash energy release and at the maximum extent of the convective zone. In model 001-7, although the entropy difference between the convective zone and the bottom of the H-rich layers is small, it is sufficient to maintain stability against convection. 

The comparison of entropy barriers among our models of a given initial composition is valid because they have identical interiors, leading to the same entropy evolution in the convective zone. The primary difference is the entropy of the envelope, explaining the trend observed in the lower panel of Fig. \ref{fig:entropy:all} of higher entropy barriers in models with higher envelope entropy. This comparison would not be valid for models with different masses or metallicities, as their He-flash evolution would differ. Indeed, \cite{2022A&A...667A.155C} noted that  it is not possible to establish a clear `entropy barrier criterion' for predicting a PIE in a given thermal pulse in AGB stars, even for models of the same metallicity. This can be understood considering that, for an AGB star,  the entropy barrier required for a PIE depends on the entropy of the H-burning shell, and the evolution of the entropy in the pulse-driven convective zone. This entropy evolution is influenced by the mass of the He-buffer and the CO-core mass of the AGB star, explaining the impossibility of finding a global critical value of the entropy barrier for the occurrence of PIEs. The metallicity dependence of the entropy barrier in our models is discussed in the next section.

\begin{figure}
	% To include a figure from a file named example.*
	% Allowable file formats are eps or ps if compiling using latex
	% or pdf, png, jpg if compiling using pdflatex
	\includegraphics[width=1\columnwidth]{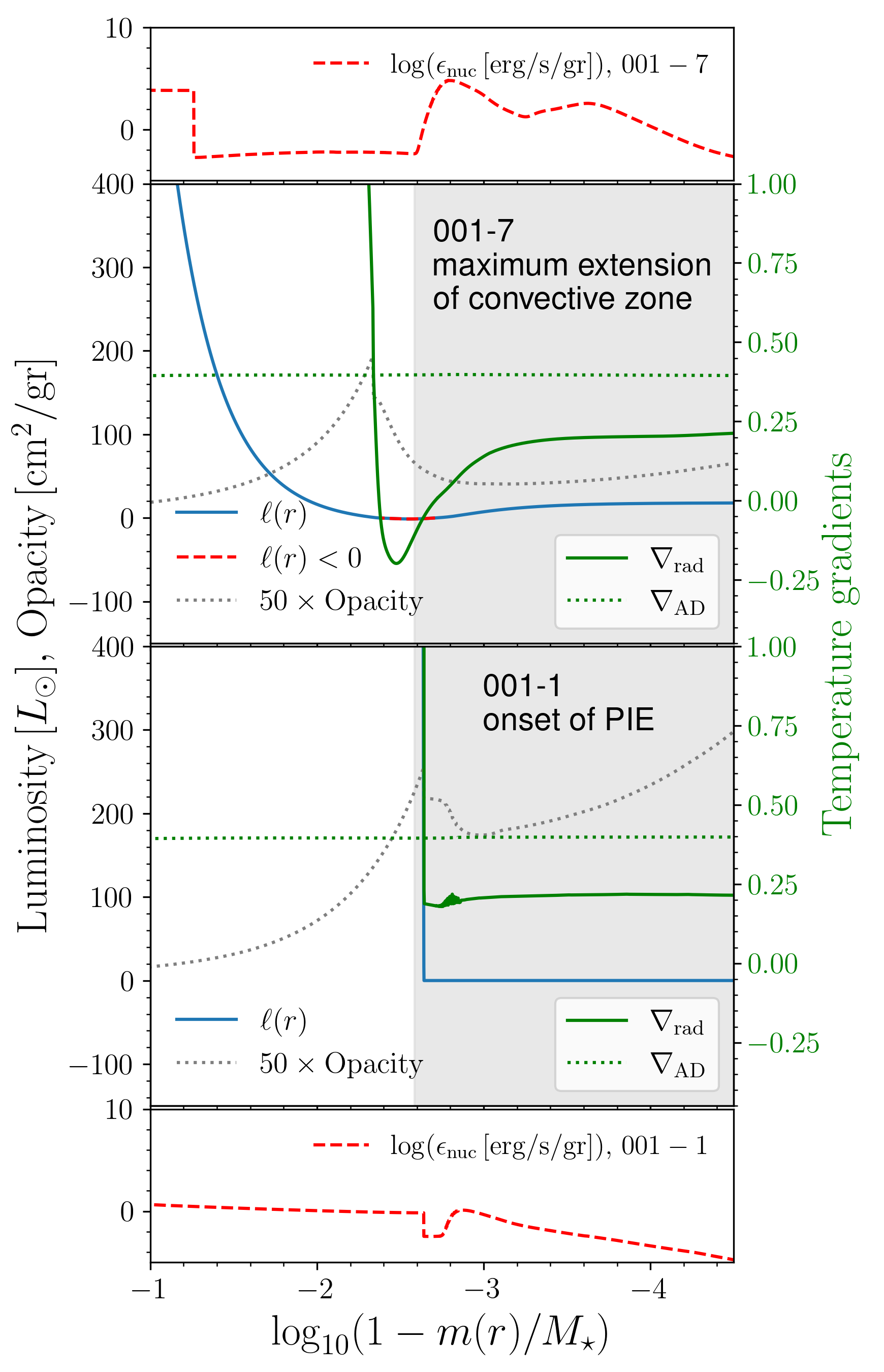}
    \caption{Temperature gradients, luminosity, opacity, and energy liberated in two of our models. The two upper panels show these quantities for model 001-7 at the moment of the maximum extension of the He-flash driven convective zone, while the two lower panels correspond to the onset of PIE in model 001-1. The red dashed line in the uppermost and lowermost panels indicate the nuclear energy liberated in each model. The run of the luminosity is shown as solid blue lines. In the 001-7 model, a red-dashed line marks where the luminosity is negative. The opacity is shown as a gray dotted line, multiplied by 50 for clarity. The radiative and adiabatic temperature gradients correspond to the solid and dotted green lines respectively, with their values shown in the right y-axis.}
    \label{fig:gradients}
\end{figure}

In Fig. \ref{fig:gradients} we compare the temperature gradients of model 001-1 and model 001-7. In model 001-7, at the moment of maximum extension of the convective zone, the material at the bottom of the H-shell burning is still stable against convection. The luminosity $\ell(r)$ here takes slightly negative values. It is worth noting that the material here is reacting to the He-flash and it is far from the steady state situation (${\rm d}S/{\rm d}t = 0$, the so-called `thermal equilibrium'). %has not yet reached a steady state again after the He-flash and is still reacting to it. 
The negative values of the luminosity run also imply negative values of the radiative gradient. In contrast, for model 001-1 at the onset of PIE, the region below the H-burning shell is unstable against convection. We note that in this model, the energy liberated via H-burning in the burning shell is $\sim 5$ orders of magnitude lower than in model 001-7.

\subsubsection{Does PIE imply i-process?}
\label{Sec:i-process}

\begin{figure}
	\includegraphics[width=1\columnwidth]{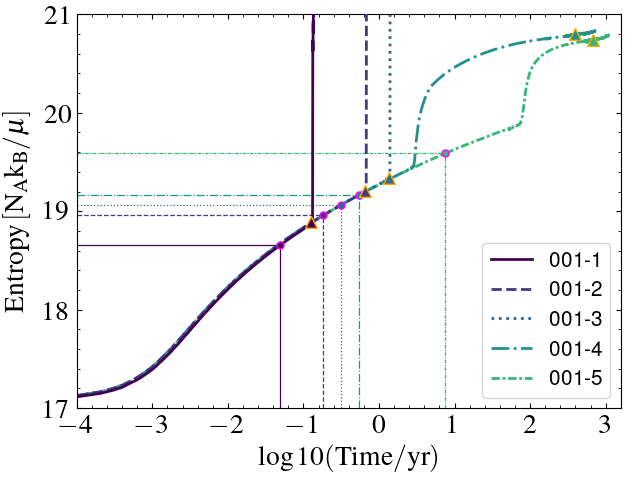}
    \caption{Evolution of the specific entropy at the He-flash convective zone of our models with initial metallicity $z=0.001$. The thin horizontal lines mark the entropy at the envelope of the models and the thin vertical lines correspond to the moment of the onset of the PIE in each model, marked also with magenta circles. The orange triangles correspond to the splitting of the convective zone in each model.}
    \label{fig:entropy:time}
\end{figure}

From different studies throughout the last decades, it has been established that the neutron densities typical of the $i$-process are attained in environments in which protons are being ingested into He-burning zones. However, it is not clear that a PIE necessarily leads to $i$-process nucleosynthesis. In our models, the occurrence of a PIE does not ensure that significant $i$-process nucleosynthesis will take place.  %From Fig. \ref{fig:abs:z0.0001-001} it is clear that . 
Only two of our five models with initial metallicity $z=0.001$ produce heavy elements in enough quantities to enrich the surface of the star (see Section \ref{sec:nuc}). 
As mentioned in Section \ref{sec:qualitative-evol-PIE}, the convective zone driven by the He-flash eventually splits into two after the onset of the PIE. Our simulations show that the models that experience significant $i$-process nucleosynthesis are the ones in which the splitting occurs hundreds of years after the onset of the PIE. These models are the ones with higher entropy barriers among those that experience PIEs (models 001-4 and 001-5, see Fig. \ref{fig:entropy:all}). 
Fig. \ref{fig:entropy:time} shows the evolution of the specific entropy for the set of models with $z=0.001$ in the convective zone driven by the He-flash. As the models are initially identical in the interior, the evolution of the He-flash and the entropy there is identical for all of them. The entropy of the material increases with time, faster at the beginning and slower when time passes. This slow-down is mostly related to the decrease in the power released by the He-flash. The outward motion of the upper convective boundary is also slowed down for the same reason.  %The development of the convective zone outwards is also slower with time. 
These models do, however, differ in the entropy of the envelope. 
The lower the entropy barrier, the earlier the entropy of the convective zone reaches the value of the specific entropy of the envelope. 
This moment is indicated in Fig. \ref{fig:entropy:time} and corresponds to the onset of the PIE in each of the models. 
The time span between the maximum of the He-flash and the onset of PIE differs by order of magnitudes for different models. For those models with higher entropy barriers, the onset of the PIE occurs when the energy liberated by the He-flash is lower and, consequently, the outer boundary of the convective zone is moving outwards at a slower pace. For example, while the convective zone of model 001-1 ingests about $\simeq 6.5\times 10^{-6}\,M_{\odot}$ of H mass in around a month, it takes 60 yr for model 001-5 to ingest the same amount. This makes the H entrainment rate lower, and therefore the burning of H is less intense. Moreover, the convective velocities are also lower. The H burning occurs mainly at the location in which the convective turnover timescale equals the timescale of the $^{12}{\rm C}+p$ reaction. Therefore, for models with higher entropy barriers, the burning of H occurs at lower temperatures, which also contributes to a less intense H burning. Consequently, in these models the drop of $\ell(r)$ at the bottom of H burning occurs at longer timescales, and therefore, the moment in which the radiative gradient equals the adiabatic one occurs also at longer timescales. This explains why the splitting of the convective zones for these models occurs orders of magnitudes later in time than in the models with lower entropy barriers. The convective splitting in each model is shown in Fig. \ref{fig:entropy:time} with triangles. For models 001-1, 001-2, and 001-3 the splitting occurs almost at the same moment of the maximum energy liberation of the H-burning. The heat added to the material by the H burning makes the entropy rise faster than it was rising due to the He flash. This can be seen happening for all of the models in Fig. \ref{fig:entropy:time}. The bulk of the H burning for models 001-1, 001-2, and 001-3, however, occurs after the splitting, and therefore, the neutron-capture processes occurring in these models before the splitting is negligible. In contrast, for models 001-4 and 00-5 significant H-burning occurs for several hundred years before the the convective zone splits. Even if the neutron densities reached here are lower than in models 001-1, 001-2, and 001-3, the amount of neutron captures occurring before the splitting is non-negligible and, as we show quantitatively in the next section, significant amounts of heavy elements are created and enrich the surface. 
All these processes can be appreciated in the different
panels of Figs. \ref{fig:KD:001-1}, \ref{fig:KD:001-4} and \ref{fig:KD:001-5}. 

The figures show how the convective zone ingests H-rich material from
the outer regions, initiating the H-flash (top two panels). This
triggers the formation and burning of $^{13}$C, leading to neutron
production (middle Kippenhahn diagrams) and ultimately the synthesis of
heavy elements, exemplified by the formation of Sr (bottom panels).
 Moreover, Figs. \ref{fig:KD:001-1}, \ref{fig:KD:001-4} and \ref{fig:KD:001-5} show that even when PIEs
happen in all cases, i-process nucleosynthesis can differ
significantly. For example, model 001-1 (Fig. \ref{fig:KD:001-1}), in which the splitting of the convective zones happens very fast, leads to a very light enrichment of heavy elements.

%Figs. \ref{fig:KD:001-1}, \ref{fig:KD:001-4} and \ref{fig:KD:001-5} show the evolution of the nuclear energy generation, the neutron density, and the abundance of Sr as a trace element in Kippenhahn diagrams for models 001-1, 001-4 and 001-5 respectively. We also show in these figures the evolution of the He and H burning luminosities, the maximum neutron density, and the maximum temperature and neutron exposures in the right y-axis, on the top panel. 

\subsection{The impact of metallicity}

\begin{figure}
	\includegraphics[width=\columnwidth]{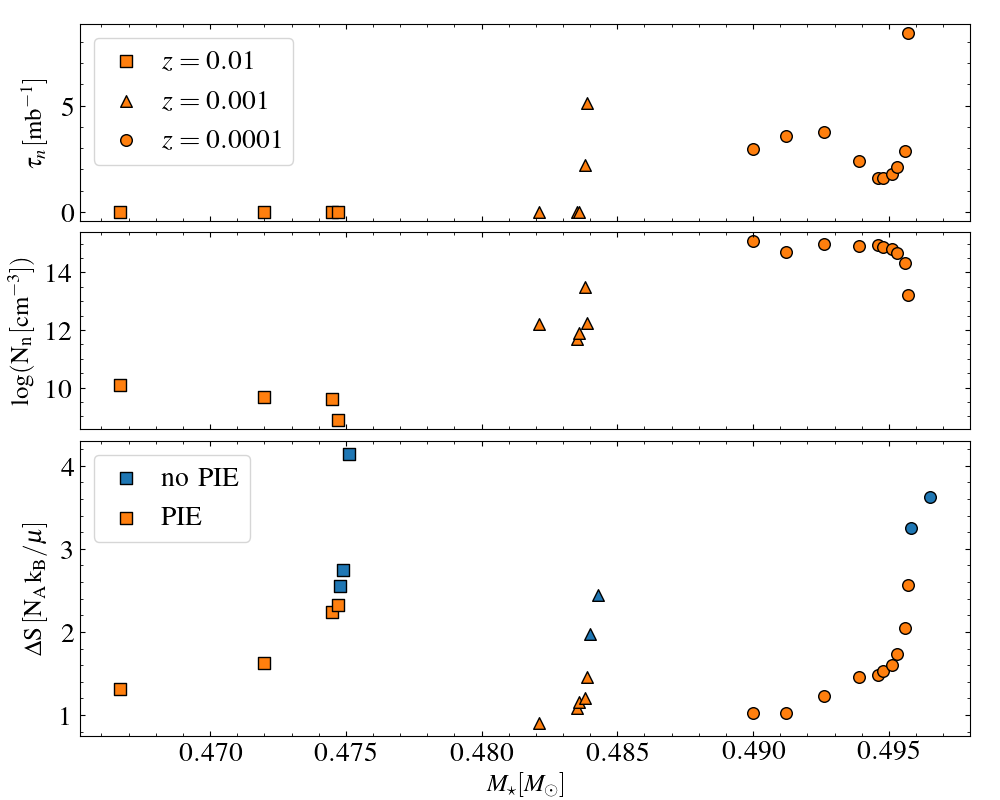}
    \caption{Maximum neutron exposure (upper panel), maximum neutron density (middle panel) and entropy barrier (lower panel) at the peak of the He flash vs the mass of each model. Blue markers  in the lower panel correspond to models that do not experience a proton ingestion episode.}
    \label{fig:all:metal}
\end{figure}

In Fig. \ref{fig:all:metal} we show the entropy barrier and the maximum values of the neutron density and the neutron exposure reached in each of the simulations of different metallicities, as a function of the mass of the models. The neutron exposure is calculated by averaging its value in the convective zone. The post-RGB masses for $z=0.01$, $z=0.001$ and $z=0.0001$ cluster around $0.47M_{\odot}$, $0.48M_{\odot}$ and $0.49M_{\odot}$ respectively.  The difference in masses for the different metallicities is due to the fact that the temperature for He burning is reached at different values of the core mass for different metallicities (see, e.g., \citealt{2017A&A...606A..33S}).
For each metallicity set (and thus almost constant core mass), the entropy
barrier height increases with envelope mass due to an increased activity
of the H-burning shell. The only exception is model 0001-1 which has a slightly higher entropy barrier than model 0001-2. However, this difference is very small (less than 1\%). The value of the entropy barrier for which a PIE is no longer possible does not have a clear dependence on the metallicity (as already pointed out for thermal pulses in AGB stars by \citealt{2022A&A...667A.155C}). In models where a PIE occurs, a higher entropy barrier (or greater model mass) corresponds to a longer time span between the onset of the PIE and the splitting of the convective zone (see Table \ref{tab:runs}). Therefore, the neutron exposure tends to be higher for models with higher mass, although it is not a monotonic relation, especially in the low-metallicity case. 
Overall, the models that are prone to having a significant production of heavy elements are those that experience a PIE but have the higher entropy barriers. One exception to this rule is model 0001-1, which experiences two H flashes. 

The values of the neutron density reached by the models change orders of magnitudes with metallicity, being higher for lower metallicities. For $z=0.01$ the values of the neutron densities are closer to the $s$-process values. In addition, the time span between the onset of PIE and the splitting of the convective zone is relatively short for the higher metallicities, implying too short neutron exposures for significant neutron captures to take place. As we discuss in the next section, we do not observe enhancement of heavy metals in the surface of any of the runs with $z=0.01$. For $z=0.001$ the neutron densities are higher, around $10^{12}-10^{13}\,$cm$^{-3}$, and we have two models with neutron exposures high enough for the production of heavy elements, as discussed in the previous section. For $z=0.0001$ the neutron densities reached can be as high as $10^{15}\,$cm$^{-3}$ and we have a higher number of models with high enough neutron exposures for nucleosynthesis to take place.

\section{Detailed nucleosynthesis}\label{sec:nuc}

\begin{figure*}
	\includegraphics[width=2\columnwidth]{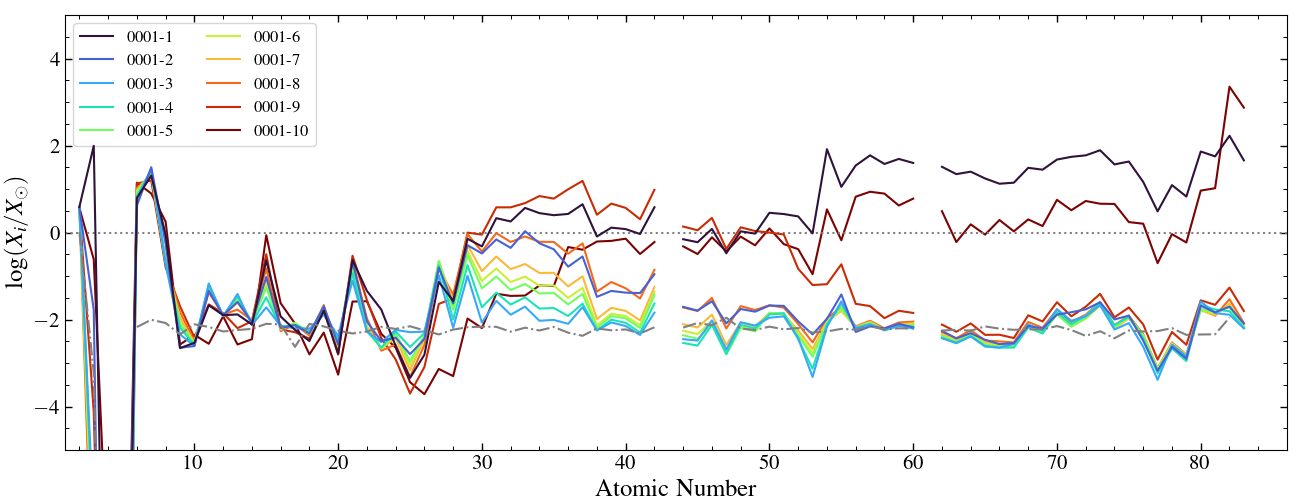}
 	\includegraphics[width=2\columnwidth]{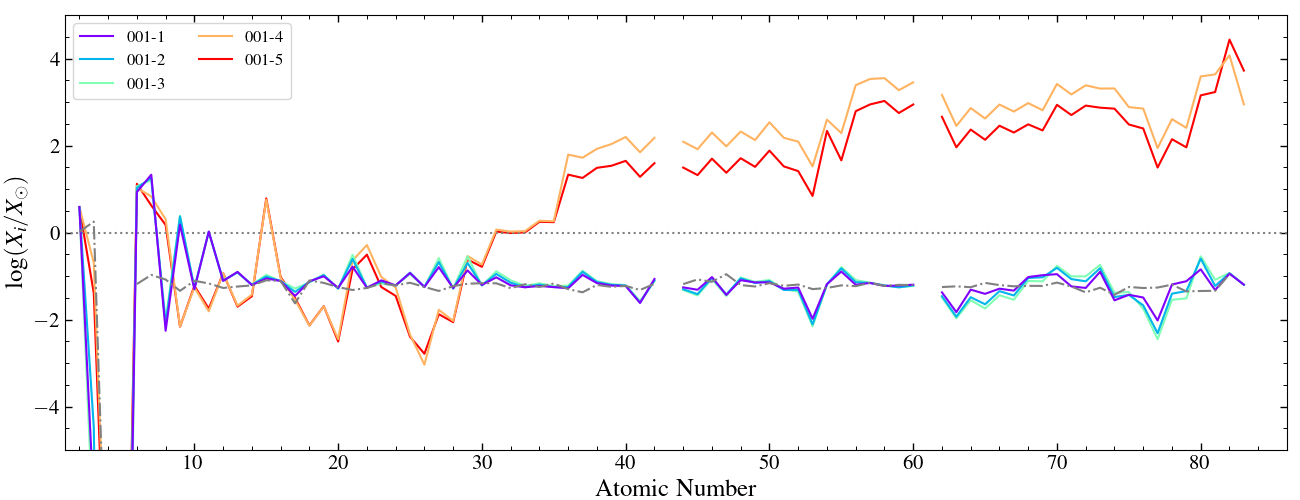}
    \caption{Superficial abundances per mass fraction respect to solar for the models with initial metallicity $z=0.0001$ (upper panel) and $Z=0.001$ (lower panel), at the zero age horizontal branch.}
    \label{fig:abs:z0.0001-001}
\end{figure*}

\begin{figure*}
	\includegraphics[width=2\columnwidth]{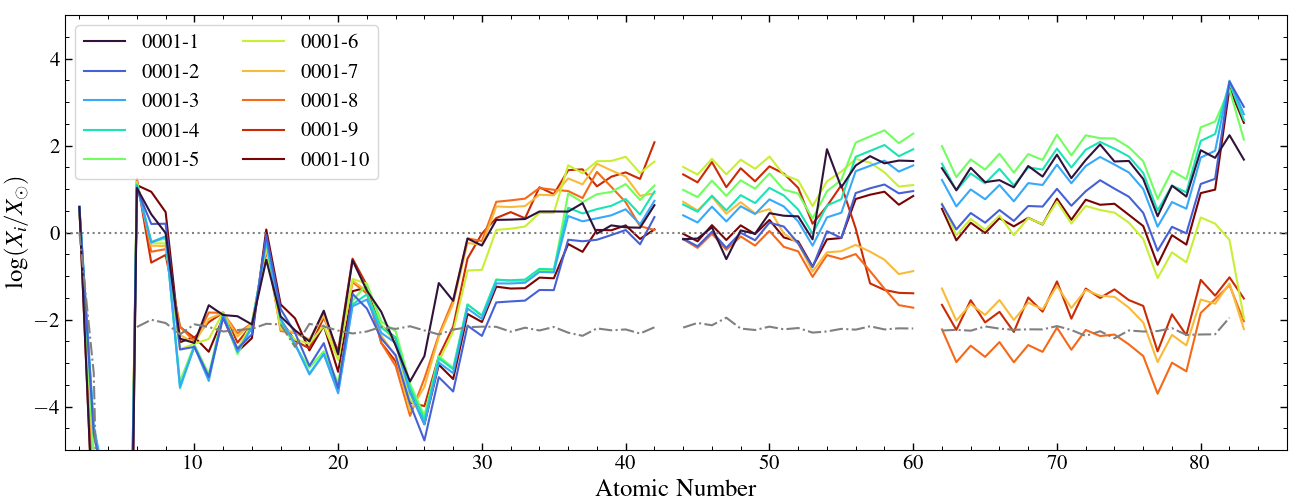}
    \caption{Abundances per mass fraction respect to solar for the models with initial metallicity $z=0.0001$, at the mass shell of the model where the maximum neutron density was reached. The abundances correspond to the last model post-processed with \texttt{ANT}.}
    \label{fig:abs:z0.0001:inside}
\end{figure*}

Figure \ref{fig:abs:z0.0001-001} shows the superficial abundances at the zero-age horizontal branch for all the runs experiencing a PIE with initial metallicities of $z=0.0001$ and $z=0.001$. The superficial abundances for $z=0.01$ are not shown  as they do not show any significant enrichment.
The initial abundances are marked with a dot-dashed gray line. For $z=0.0001$, runs 0001-1 and 0001-10 show abundance enhancements of all heavy elements for $Z>30$. Abundance enhancements of these elements are of one to four orders of magnitudes and tend to increase with $Z$. 
Model 0001-10 (brown line in the upper panel of Fig. \ref{fig:abs:z0.0001-001}) is the one which experiences a longer time between the onset of the PIE and the splitting of the convective zone (see  Table \ref{tab:runs}). In this case, we obtain a marked increase in Pb ($Z=82$) abundance of more than 4 orders of magnitude. Elements between $Z\simeq 53$ and $Z=81$ show abundance increases of 2 to 2.5 orders of magnitude. Run 0001-1 (black line in the upper panel of Fig. \ref{fig:abs:z0.0001-001}) 
 is a particular model that experiences two H flashes (see Fig. \ref{fig:two:flashes}). This holds true when we change the spatial and temporal resolution of the models, pointing to a physical reason behind these two flashes. After the first flash the convective zone splits into two, but about one month later the convective zones reconnect again and the second flash occurs. This  model presents significant enrichment of elements for atomic numbers $Z\sim30$ to $Z\sim54$ of about 2 dex, and for elements beyond $Z\simeq 54$  of about 3 to 4 dex. From run 0001-3 to 0001-8 we observe a monotonic increase in the surface enrichment of heavy elements.  The most enriched models are the ones with longer intervals between the onset of the PIE and the splitting of the convective zone. Sequences up to 0001-7 show enrichment in elements up to $Z\simeq 40$ but not beyond, while model 0001-8 also presents enrichment of elements up to $Z\simeq 60$. The exception to this monotonic behavior is model 0001-2.  The timespan in which the H-burning luminosity is higher than $\log(L_{\rm CNO}/L_{\sun})\sim8$ is around twice for this model when compared to 0001-3. Therefore, even if the time between the onset of PIE and the splitting of the convective zone is similar for both models, 0001-2 and 0001-3, the neutron exposure before the splitting is higher for model 0001-2. This is not reflected on the neutron exposures listed in Table \ref{tab:runs} as they correspond to the maximum value obtained during the whole evolution, even after the splitting of the convective zones. For $z=0.001$ we only obtain two extreme cases: either the models do not show any enrichment in heavy elements, or they show high enhancements in all heavy elements beyond $Z=30$ with an abundance pattern similar to the one obtained for 0001-10. Both models 001-4 and 001-5 present very high abundances of elements beyond $Z=54$, enhanced in 3 to 5 dex, and a significant peak in Pb.

In Fig. \ref{fig:abs:z0.0001:inside} we show the interior abundances for $z=0.0001$, in the zone where the maximum neutron exposure is reached. For each run, the abundances plotted correspond to the last model for which the abundances have been calculated via post-processing. 
Comparing with the upper panel of Fig. \ref{fig:abs:z0.0001-001} we see that neutron capture processes continue happening inside the models after the splitting of the convective zone. For six of the sequences, the abundances of heavier elements continue rising leading to a peak in Pb. If any type of extramixing or CBM process is able to reconnect the outer convective zone and the inner He-flash driven convective zones this would lead to a further enhancement of very heavy elements at the surface (see Section \ref{conclusion}).

\section{Discussion}\label{sec:discuss}
\label{conclusion}

\subsection{Current limitations of our models}
\label{limitations}
As discussed in previous sections, significant $n$-capture nucleosynthesis happens for a minority of our simulations. Most of the sequences that experience significant $n$-capture nucleosynthesis (001-5, 001-4, 0001-9 and 0001-10) correspond to the largest masses that experience a PIE for each metallicity. These sequences are the ones close to the threshold envelope masses for which the H-burning shell prevents the development of a PIE (see Table \ref{tab:runs}). This is not a coincidence as for these sequences the relatively high entropy barrier is able to delay the development of the violent H-burning runaway that leads to the splitting of the convective zones. The only exception to this trend is the other threshold sequence, 0001-1, which corresponds to the lowest mass that experiences a He-core flash at $z=0.0001$. 
We see that only models  in a small mass interval $\Delta M^{\rm nuc}_\star\simeq 0.0001$ to $0.0002 M_\odot$ experience significant nucleosynthesis, out of a much larger mass interval of all the models that experience a PIE ($\Delta M^{\rm PIE}_\star\simeq 0.002$ to $0.005 M_\odot$, see Tab. \ref{tab:runs}). If no fine-tuning of mass loss is assumed this suggests that only a small fraction of about $\sim 2$\% ($Z=0.0001$) to $\sim 10$\%   ($Z=0.001$) of all He-rich subdwarfs formed through this channel would become enriched in neutron capture elements. As discussed in Section \ref{Sec:i-process} the $i$-process nucleosynthesis in the models is very sensitive to the interval between the maximum of the He-flash and the splitting of the convective zones.
Consequently, any physics neglected in the current models able to hinder the development of the H-flash might favour the formation of heavy metals in H-deficient subdwarf stars. The most obvious  candidates are chemical gradients ($\nabla \mu$). In the current models convective instability (i.e. Schwarzschild criterion) and mixing velocities were computed neglecting the impact of $\nabla \mu$. The strong chemical gradient between the core and the H-rich envelope is expected to make convective instability more difficult (i.e. Ledoux criterion) and mixing slower \citep[see][]{2006A&A...449..313M}. Both effects should delay the splitting of the convective zone and favour $i$-process nucleosynthesis. Moreover, the development of the H-flash happens extremely fast, not far from the convective turnover timescale. Under these conditions the instantaneous adjustment of temperature gradients and velocities assumed in MLT might not be a good approximation \citep{2002ASPC..259..456F,2022A&A...667A..97A}. If this is so, it is also possible that the inertia of the convective motions might keep H-burning and He-burning regions coupled for a longer time, favoring $i$-processes.  Moreover, any CBM process able to mix part of the inner and outer convective zones after the splitting would favor the formation of He-rich subdwarfs with enhanced heavy elements. In this context, it is interesting to note that 3D hydro models suggest that the first stages of PIE events might not be as steady as they are in 1D stellar models. \cite{2011ApJ...727...89H} find in their 3D-hydro models that the early stages of H-entrainment are inhomogeneous, asymmetric,
as well as intermittent, allowing H to be advected into the deeper
layers where the burning will eventually take place. Moreover,  \cite{2014ApJ...792L...3H} find in
some of their simulations the development of global non-spherical
oscillations sustained by ignition events of H-rich pockets. Maybe even more relevant for the formation of hot-subdwarf stars are the results from the 3D-hydro models of H-ingestion flashes after the He-core flash. In their simulations, \cite{2011A&A...533A..53M,2012ASSP...26...87M} find that the early stages of the PIE develop slightly differently than in 1D stellar models. They find that the temperature peak created by H-burning leads first to a retreat of the outer boundary of the He-burning convective zone, and later to the appearance of a new H-burning shell. Most importantly, \cite{2011A&A...533A..53M,2012ASSP...26...87M} found that in the early stages of the PIE, there is an absence of an ``impermeable'' radiative layer between the two convective zones. Consequently, in the early stages of a PIE probed by these computations, the mixing of nuclear species
between these zones keeps occurring during the
whole simulations (with decreasing efficiency towards the end
of the computation).

 Another point of concern is the occurrence of CBM during the PIE and its impact on the final heavy metal abundances \citep{2024A&A...687A.260R, 2024A&A...684A.206C}. Unfortunately, all proposed CBM prescriptions in use in stellar evolution codes are oblivious to the presence of chemical gradients or the entropy barrier. Consequently, a quantitative estimation of the impact of such processes is not feasible at the moment. Recently \cite{2024A&A...684A.206C} performed a parameter exploration of the the impact of CBM on i-process nucleosynthesis in AGB models. Their results hint that CBM, if it exists, might be important at the upper boundary of the He-flash driven convective zone. They find that variations of CBM efficiencies at the top of the helium flash convective zone leads to a scatter of 0.5 to 1 dex in the abundances of elements with $36 < Z < 56$, and almost no impact on the abundances of elements with $56 < Z < 80$.
We plan to address the impact of CBM, time-dependent convection, and chemical gradients in future works.

 Finally, nuclear model uncertainties also have an impact on the predictions of i-process nucleosynthesis computations \citep{2018JPhG...45e5203D, 2024A&A...684A...8M}. Specifically, \cite{2024A&A...684A...8M} ﬁnd that these uncertainties have a significant impact of 0.5 to 1 dex in the abundances of elements with $Z\gtrsim 40$, which is actually the range in which our models fail to reproduce the abundances observed in EC\,22536-5304 (see next section).

\begin{figure*}
\includegraphics[width=\textwidth]{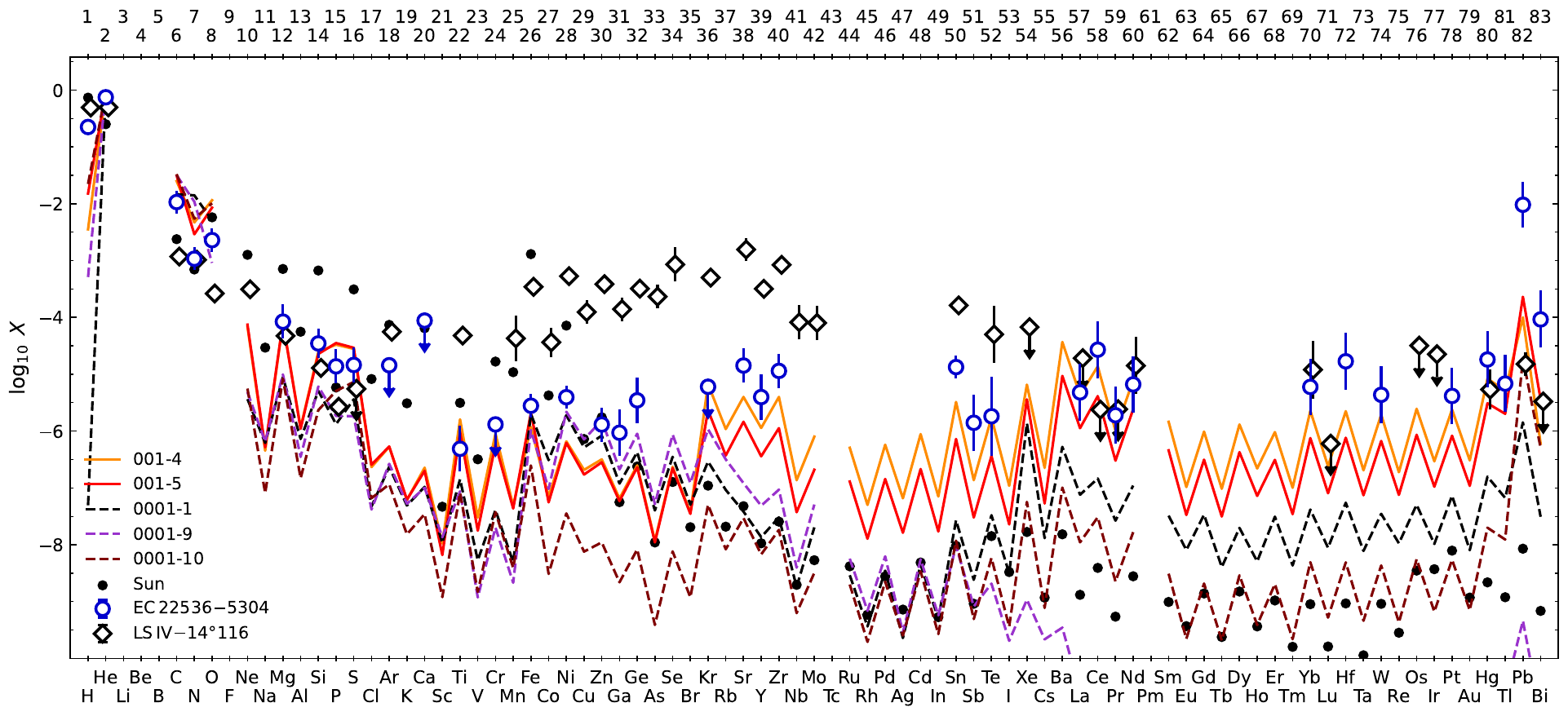}\\
\caption{Abundances by mass fraction $X$ compared to observations of two helium-rich hot subdwarf stars LS\,IV\,$-$14\,116 \citep{2020A&A...643A..22D}, EC\,22536$-$5304 \citep{2021A&A...653A.120D}. These stars have two different suspected evolutionary origins (see text). Abundances of LS\,IV\,$-$14\,116 and EC\,22536$-$5304 include preliminary values obtained from HST far-UV spectra \citep{2022hst..prop17072D}, which will be described in detail in a future paper. 
Solar abundances from \cite{2009ARA&A..47..481A} and updated by \cite{2015A&A...573A..27G} are shown for comparison. 
}
\label{fig:stars}
\end{figure*}

\begin{figure*}
\includegraphics[width=\textwidth]{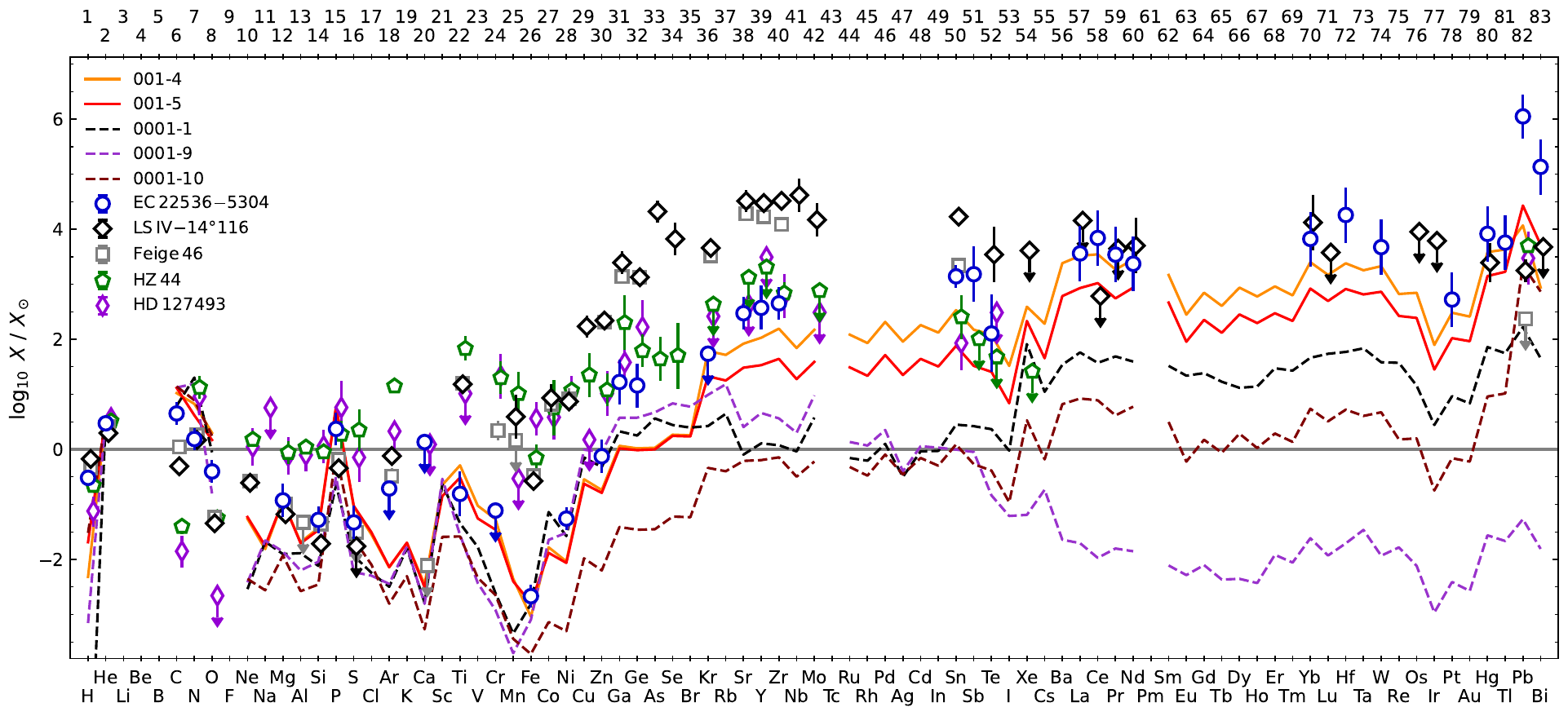}\\[0pt]
\includegraphics[width=\textwidth]{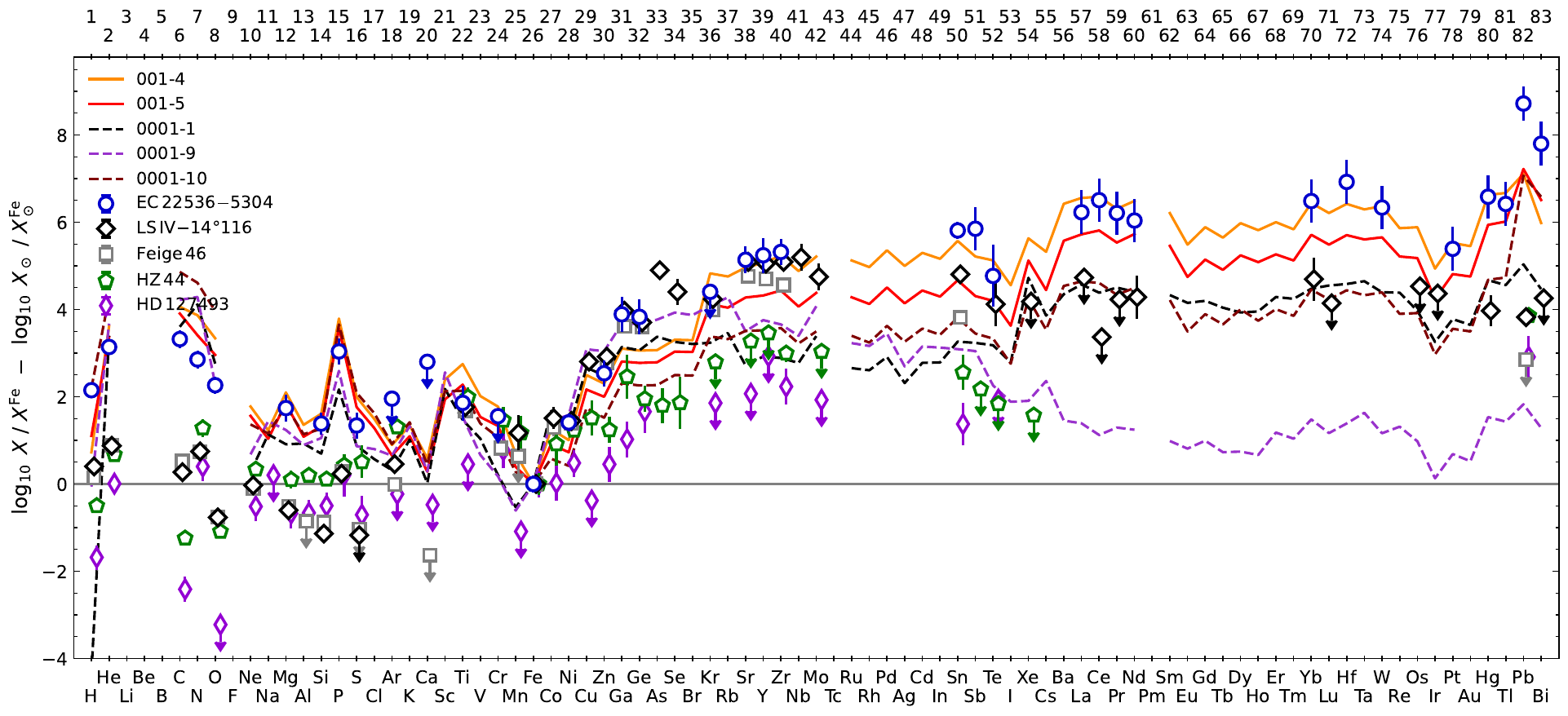}
\caption{Upper panel:  Surface abundances by mass fraction $X$ (relative to solar) of our models with significant neutron capture nucleosynthesis compared to observations of the helium-rich hot subdwarf stars LS\,IV\,$-$14\,116 \citep{2020A&A...643A..22D}, EC\,22536$-$5304 \citep{2021A&A...653A.120D}, Feige\,46 \citep{2019A&A...629A.148L}, HZ\,44 and HD127493 \citep{2019A&A...630A.130D}. Lower panel: Same as the upper panel by showing the abundances relative to iron, which form the main nuclei seed of the synthesized heavy elements. The abundances of LS\,IV\,$-$14\,116 and EC\,22536$-$5304 include preliminary values obtained from HST far-UV spectra \citep{2022hst..prop17072D}, which will be described in detail in a future paper. }
\label{fig:stars2}
\end{figure*}

\subsection{Comparison with observations}

Among all the heavy metal He-sdOB stars, EC 22536-5304 offers a unique opportunity for  testing our models. As mentioned before, this star is a member of a binary system. Its sdF companion allows for the determination of the initial metallicity of the system \citep[$\mathrm{[Fe/H]}\approx-2$, ][]{2021A&A...653A.120D}. Moreover, the 
current orbital configuration of the system ($P=457$d) indicates that the system underwent a  Roche lobe overflow (RLOF) evolution that led to the formation of the hot-subdwarf. This fact, together with the high He content of EC 22536-5304 points at a hot-flasher origin for this star. Fig. \ref{fig:stars} compares the absolute abundances of our models with significant neutron-capture nucleosynthesis with those of EC 22536-5304. The agreement with the predicted $i$-process abundances in our models up to $Z=60$ is remarkable.
 These results support both our hypothesis of a hot-flasher origin for EC~22536-5304 and the occurrence of i-process nucleosynthesis, as predicted by our computations. The inability of our models to quantitatively match the observed abundances of very heavy elements ($Z>60$) might hint at the limitations in our current stellar models, particularly in the treatment of convection (see Section \ref{limitations}), as well as the absence of diffusion processes. Figure \ref{fig:stars} also compares the abundances of our model with another very well-studied He-sdOB, LS IV-14$^\circ$116. 
LS IV-14$^\circ$ 116 and its twin Feige 46  (see Fig. \ref{fig:stars2}) are two pulsating He-sdOBs for which detailed abundances pattern have been determined. 
In addition, these stars show halo kinematics  \citep{2019A&A...629A.148L, 2021A&A...653A.120D}, hinting at a low initial metallicity. Clearly, the abundance trend of these stars for intermediate heavy elements ($18\lesssim Z\lesssim 52$) is notably different from those of our simulations. Notably, the best evolutionary scenario for these two stars is not the hot-flasher scenario. \cite{2022MNRAS.511L..60M} demonstrated that both their pulsation properties and surface composition can be simultaneously explained by the rare merger of a He-WD with a low-mass CO-WD, or the merger of two low-mass CO-WDs, which are expected to feature large He envelopes \citep{2019MNRAS.482.1135Z}. 
 Overall, these results indicate that the star-to-star differences in heavy
metal abundances can indeed be used to identify their evolutionary origin.

%Overall, these results not only indicate that hot-flasher models do a good job at explaining the abundances of some heavy metal He-sdOB stars but also that the star-to-star differences in heavy metal abundances can indeed be used to identify their evolutionary origin.

In Fig. \ref{fig:stars2} we compare the abundances of the models with significant $i$-process nucleosynthesis with those of five very well studied heavy-metal iHe-sdOBs \citep{2019A&A...630A.130D,2021A&A...653A.120D, 2020A&A...643A..22D}, including EC 22536-5304 and LS IV-14$^\circ$116. Abundances in the upper panel plot are shown in mass fraction relative to solar. 
This compares the nucleosynthesis yields of our models with the observed stellar abundances.
The abundances in the lower panel are expressed as mass fractions relative to the star’s iron content and to solar values ($[X_i/{\rm Fe}] = \log(X_i /X_{\rm Fe} )-\log(X_i /X_{\rm Fe} )|\odot $). This highlights the abundance pattern of the produced trans-iron elements relative to the seeds of the neutron-capture processes in the iron group.
 Notably, the trend in the relative abundances (lower panel of Fig.  \ref{fig:stars2}) predicted by the $i$-process nucleosynthesis between $Z=26$ (iron) and $Z=42$ (molybdenum) covers the values observed in well studied heavy-metal hot subdwarfs, hinting at a possible similar $i$-process origin. No model perfectly reproduces any single star, a fact very likely related to the limitations both in the treatment of convection (Section \ref{limitations}) and the absence of radiative levitation, as well as the possibility of a different evolutionary origin for some of these stars (e.g. LS\,IV\,$-$14\,116 and Feige\,46, see previous paragraph, but also HZ44 and HD127493).

Abundances with respect to the solar abundances (Fig.\ \ref{fig:stars2}, upper panel) show an interesting but diverse landscape. 
As mentioned above, the abundances of the most likely hot-flasher star (EC 22536-5304) are well reproduced by our models up to lead, which has higher abundance in the observation. 
%$Z= 50$ (tin, Sn).
This is true not only for trans-iron elements but also for intermediate and light elements. This might be an indication that EC 22536-5304 has experienced its hot-flasher event recently enough so that radiative levitation has not yet been able to alter significantly the abundances pattern coming from the nucleosynthesis. Trans-iron elements below lead in EC 22536-5304 are remarkably well reproduced by our models with the higher heavy element abundances (001-4 and 001-5), 
while lead and bismuth are about one order of magnitude higher in EC 22536-5304 than in the simulations. Conversely, while the absolute abundances of  LS\,IV\,-14$^\circ$116 and Feige\,46 between argon (Ar, $Z=18$) and tellurium (Te, $Z=52$)
 are about two orders of magnitude larger than in the simulations, heavier element abundances ($Z>52$) in both stars are well reproduced by the models. 
Abundances in HZ44 and HD127493 are different from the previous two cases. 
Abundances of the heaviest measured trans-iron element (Sn) are very similar to the models, but differences increase as we move to lighter elements. Light trans-iron elements (Zn, Ga, Ge, As, Se) are about one order of magnitude higher than our models, iron-group elements are about two orders of magnitude higher than the models and differences are even larger for intermediate elements (Ti, Ar, S) and light elements such as Ne, Al, and Mg. 
Given their higher light-elements abundances, it seems likely that HZ44 and HD127493 have higher initial metallicities than the other three stars. This is consistent with their thin (HD127493) and thick disk (HZ44) kinematics \citep{2021A&A...653A.120D}. 

The five studied stars can be separated into three different groups: EC~22536-5304 with high relative abundances between iron and molybdenum and very high relative abundances between tin and lead, LS\,IV\,-14$^\circ$116 and Feige\,46 with high element abundances  between iron and molybdenum and high relative abundances between tin and lead (but two orders of magnitude lower than the previous group), and HZ44 and HD127493  which show moderately high element abundances between iron and barium (about two orders of magnitude below the previous group) and lead abundances similar to the second group. Notably, if we separate these five stars by their most likely evolutionary scenario according to independent observables we get the same groups: the binary nature of EC~22536-5304 coupled with its high surface helium abundance strongly suggests a hot-flasher event after a RLOF stage \citep{2021A&A...653A.120D}, the nature of the $g$-mode pulsation and surface properties of LS\,IV\,-14$^\circ$116 and Feige\,46  have only been reproduced by mergers involving very low-mass CO-WDs \citep{2022MNRAS.511L..60M}, and finally, the distinctive CNO-cycle pattern (N-rich,
C/O-poor) observed in HZ44 and HD127493 hints at the slow merger of He-WDs \citep{2012MNRAS.419..452Z, 2018A&A...620A..36S}. This might be an indication that different evolutionary channels lead to different trans-iron element patterns and that radiative levitation processes are not efficient enough to remove these differences. 

In light of the qualitative agreement of the abundances relative to iron between $i$-process nucleosynthesis computations and iHe-sdOB stars, a self-synthesized origin enhanced by diffusion processes appears plausible. This idea is supported by the possibility of PIEs during merger events, also leading to $i$-process nucleosynthesis. A proper assessment of the relevance of radiative levitation requires the computation of time-dependent radiative levitation for trans-iron elements. Such simulations are yet to be performed. \cite{2011A&A...529A..60M} carried out time-dependent radiative levitation calculations in models of hot-subdwarf stars but only for elements up to Ni, and only for helium-poor stars. Recently, \cite{2024MNRAS.530.2039S} investigated how lead stratification in the atmosphere affects the spectra of hot subdwarf stars. 
They argue that to explain the observed lead abundances by radiative levitation alone, and assuming a solar initial composition, radiative levitation should concentrate all the lead available in $10^{-9}\,M_{\odot}$ (up to $10^{-5}\,M_{\odot}$ for the extreme case of EC 22536-5304) in the line-formation region. It is yet to be proven whether radiative levitation can achieve this.

\section{Conclusions}\label{sec:conclusions}
We have computed the neutron-capture nucleosynthesis in models of hot-subdwarf stars constructed by stripping RGB stellar models, for different stellar metallicities (Z=0.01, 0.001, 0.0001). Our simulations show that  proton ingestion episodes (PIE) can lead to $i$-process nucleosynthesis in models with $Z\leq 0.001$. We find that not all PIEs lead to $i$-process nucleosynthesis. Specifically, we find that significant $i$-process nucleosynthesis only develops for lower metallicity models ($Z=0.001$, $0.0001$) and for those models that have the longest intervals between the main He-core flash and the PIE. Within the standard `MLT+Schwarzschild Criterion' picture this corresponds to the models with the most active H-burning shell that still allows the development of the PIE. 
We also find significant $i$-process nucleosynthesis in a model that barely develops the He-core flash and shows two H flashes. 
Our analysis shows that the relatively high specific entropy difference between the core and the H-rich material is able to delay the runway ingestion of protons, giving more time for the neutron capture nucleosynthesis to develop before the splitting of the convective zones. Thus, under the standard MLT-based mixing and burning picture some fine-tuning is required to obtain significant trans-iron elements production. Any physical process able to delay the splitting of the convective zones, or to keep them connected, would ease the synthesis of trans-iron elements. Future works will explore the impact of chemical gradients and CBM processes in the $i$-process nucleosynthesis.

Comparing the abundances from our $i$-process nucleosynthesis calculations with those of iHe-sdOB stars provides an interesting perspective. Overall, the heavy element abundances relative to iron show qualitative agreement with the $i$-process models, suggesting that PIEs and $i$-process nucleosynthesis may have occurred during the formation of these stars. Absolute abundances of trans-iron elements, however, are generally lower in our models than in the five stars that have been studied in greatest detail. This might be an indication that iHe-sdOB stars  display the effects of $i$-process nucleosynthesis enhanced by radiative levitation.

Perhaps more interesting is the fact that for EC 22536-5304, the element abundances up to tin closely match the nucleosynthesis predictions of our models. EC 22536-5304 is the only star among the five studied for which the hot-flasher scenario, as assumed in our models, is the most likely evolutionary path. 
 This agreement supports the hypothesis that EC 22536-5304 underwent a hot-flasher event following a stage of RLOF, and show the potential of this scenario for i-process nucleosynthesis.

The other four studied stars also show trans-iron element abundances that correlate with their suspected origins: LS\,IV\,-14$^\circ$116 and Feige\,46 
\citep[likely produced by mergers involving at least one low-mass CO-WD,][]{2022MNRAS.511L..60M} have significantly different abundances from those of HZ44 and HD127493 \citep[likely produced by the slow merger of He-WDs][]{2012MNRAS.419..452Z, 2018A&A...620A..36S}, and also from those of the likely hot-flasher EC 22536-5304.
If this separation in groups is confirmed, it would indicate that diffusion processes are not the sole explanation of observed abundances in iHe-sdOB stars, but that their formation channel is also relevant. 
Moreover, if the formation scenario indeed determines the observed abundance pattern of iHe-sdOBs, then we could use heavy-element measurements for a sample of these stars to constrain the relative frequency of each formation channel.

\begin{acknowledgements}
We thank the referee for the constructive report, which helped to improve this work. T.B. thanks S.E. de Mink for useful discussions regarding the presentation of this work. Part of this work was supported by the German \emph{Deut\-sche For\-schungs\-ge\-mein\-schaft, DFG\/} project number Ts~17/2--1.This research was supported by the Munich Institute for Astro-, Particle and BioPhysics (MIAPbP) which is funded by the Deutsche Forschungsgemeinschaft (DFG, German Research Foundation) under Germany´s Excellence Strategy – EXC-2094 – 390783311. Part of this work was supported by the CONICET-DAAD 2022  bilateral cooperation grant number 80726. M3B is partially funded by CONICET and Agencia I+D+i through grants PIP-2971 and PICT 2020-03316. Matti Dorsch is supported by the Deutsches Zentrum für Luft- und Raumfahrt (DLR) through grant 50-OR-2304. A.S. acknowledges support by the Spanish Ministry of Science, Innovation and Universities through the grant PID2023-149918NB-I00 and the program Unidad de Excelencia Mar\'{i}a de Maeztu CEX2020-001058-M, and by Generalitat de Catalunya through grant 2021-SGR-1526.
\end{acknowledgements}

% WARNING
%-------------------------------------------------------------------
% Please note that we have included the references to the file aa.dem in
% order to compile it, but we ask you to:
%
% - use BibTeX with the regular commands:
%   \bibliographystyle{aa} % style aa.bst
%   \bibliography{Yourfile} % your references Yourfile.bib
%
% - join the .bib files when you upload your source files
%-------------------------------------------------------------------

\bibliographystyle{aa} % style aa.bst
\bibliography{bibliografia}

\begin{appendix}
\section{Neutron exposure of our models}\label{appx:1}

The neutron exposure is:
\begin{equation}
     \tau_{n} = v_{T}\int n_n(t)\, dt
\end{equation}
where $n_n$ is the number density of neutrons and $v_T$ their thermal velocity,
\begin{equation}
    v_T = \left(\frac{2kT}{\mu_n} \right),
\end{equation}
being $\mu_n \simeq m_n$, the mass of a neutron. The material that has been exposed to neutrons is being mixed. Therefore, we `mix' the neutron exposure with a similar approach as the scheme of \cite{2001ApJ...554.1159C} for the abundances:
\begin{equation}
    \tau_{n,i} =\,^0\!\tau_{n,i} + \frac{1}{V_{\rm conv}} \sum_{j = {\rm conv}} (^0\!\tau_{n,j}-\,^0\!\tau_{n,i})f_{ij} \Delta V_j,\label{eq:mix-neuxp}
\end{equation}
where $\tau_{n,j}$ is the (total) neutron exposure of shell $j$, $\Delta V_j$ is the volume of shell $j$, $V_{\rm conv}$ the volume of the convective region and $f_{ij}$ is the same damping factor defined by \cite{2001ApJ...554.1159C}. 
The quantity $^0\!\tau_{n,j}$ is constructed as follow:
\begin{align}
    ^0\!\tau_{n,j} = \tau_{n,j}\,(\text{from previous time step}) + {\rm d}\tau_{n,j}\,,
\end{align}
where ${\rm d}\tau_{n,j} = v_{T,j}\,\,n_{n,j}$ for the current time step. The motivation for using the volume instead of the mass is the definition of the neutron exposure, which depends on the number density of neutrons, instead of on the mass fraction. This might not be strictly correct, but we are interested mainly in the order of magnitude of the neutron exposure.

\section{Evolution of model 0001-1}\label{appx:2}

\begin{figure}
	\includegraphics[width=1\columnwidth]{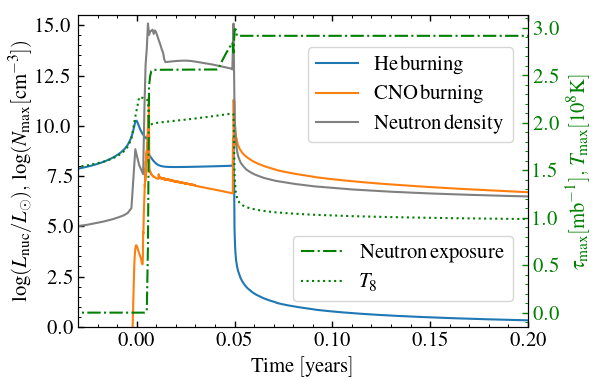}
    \caption{Evolution of the He burning luminosity (blue line), the CNO burning luminosity (orange line), the neutron density (gray line) for model 0001-1. The right y-axis corresponds to the neutron exposure, shown in a dot-dashed green line, and the maximum temperature of the model at each time, shown in a dotted green line.}
    \label{fig:two:flashes}
\end{figure}

\end{appendix}

\end{document}